\documentclass{article}

\usepackage{amsmath}
\usepackage{amsfonts}
\usepackage{amssymb}
\usepackage{graphicx}
\usepackage{color}

\usepackage{cite}
\usepackage{bm}

\usepackage[numbers]{natbib} 

\usepackage{enumitem}
\usepackage[margin=.75in]{geometry}
\usepackage{url}

\usepackage[normalem]{ulem}
\usepackage{xcolor} 

\definecolor{dartgreen}{RGB}{0, 105, 62}

\newcommand{\sgn}{\mathop{\mathrm{sgn}}}

\title{Energy landscape analysis based on the Ising model: Tutorial review}
\author{Naoki Masuda$^{1,2,*}$, Saiful Islam$^{2}$, Si Thu Aung$^{1}$, and Takamitsu Watanabe$^{3}$\\ \ \\
$^1$ Department of Mathematics, State University of New York at Buffalo,\\
Buffalo, New York, United States of America\\
$^2$ Institute for Artificial Intelligence and Data Science, State University of New York at Buffalo,\\
Buffalo, New York, United States of America\\
$^3$ International Research Centre for Neurointelligence, The University of Tokyo, Tokyo, Japan\\
$^*$ Corresponding author (naokimas@gmail.com)}
\date{}

\begin{document}
\maketitle

\begin{abstract}
We review a class of energy landscape analysis method that uses the Ising model and takes multivariate time series data as input. The method allows one to capture dynamics of the data as trajectories of a ball from one basin to a different basin to yet another, constrained on the energy landscape specified by the estimated Ising model. While this energy landscape analysis has mostly been applied to functional magnetic resonance imaging (fMRI) data from the brain for historical reasons, there are emerging applications outside fMRI data and neuroscience. To inform such applications in various research fields, this review paper provides a detailed tutorial on each step of the analysis, terminologies, concepts underlying the method, and validation, as well as recent developments of extended and related methods.
\end{abstract}


\section{Introduction\label{sec:introduction}}

\begin{figure}
\begin{center}
\includegraphics[width=15cm]{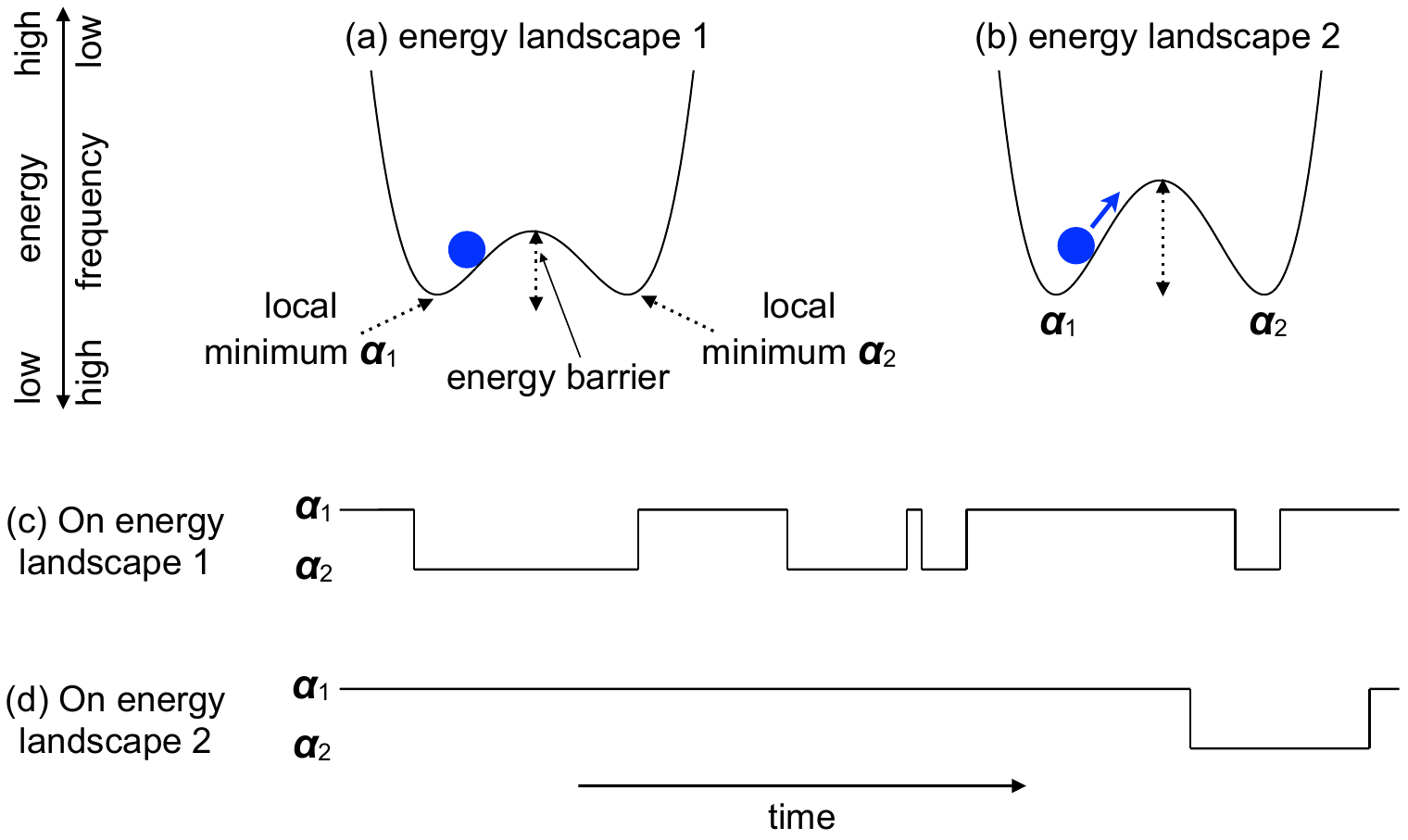}
\caption{One-dimensional schematic of energy landscapes. (a) An energy landscape with two local minima of energy, $\bm{\alpha}_1$ and $\bm{\alpha}_2$. They are separated by an energy barrier. The filled circle, or ``ball'', represents the state of the system, which tends to go downhill but can also go uphill with a smaller rate.
The position of the ball along the horizontal axis specifies the system's state, which we call the activity pattern. (b) An energy landscape with a higher energy barrier between two local minima. (c) Sample time course of the system's dynamics on the energy landscape shown in (a). (d) Sample time course on the energy landscape shown in (b). Note that switching between $\bm{\alpha}_1$ and $\bm{\alpha}_2$ is rarer in (d) than (c) because $\bm{\alpha}_1$ and $\bm{\alpha}_2$ are separated by a higher energy barrier in (b) than (a). In fact, the system's state (i.e., filled circle) can take activity patterns different from the two local minima (e.g., a pattern close to but different from $\bm{\alpha}_1$). However, we only show stochastic transition dynamics between $\bm{\alpha}_1$ and $\bm{\alpha}_2$ in (c) and (d) for simplicity.}
\label{fig:1dim-schem}
\end{center}
\end{figure}

\begin{figure}
\begin{center}
\includegraphics[width=15cm]{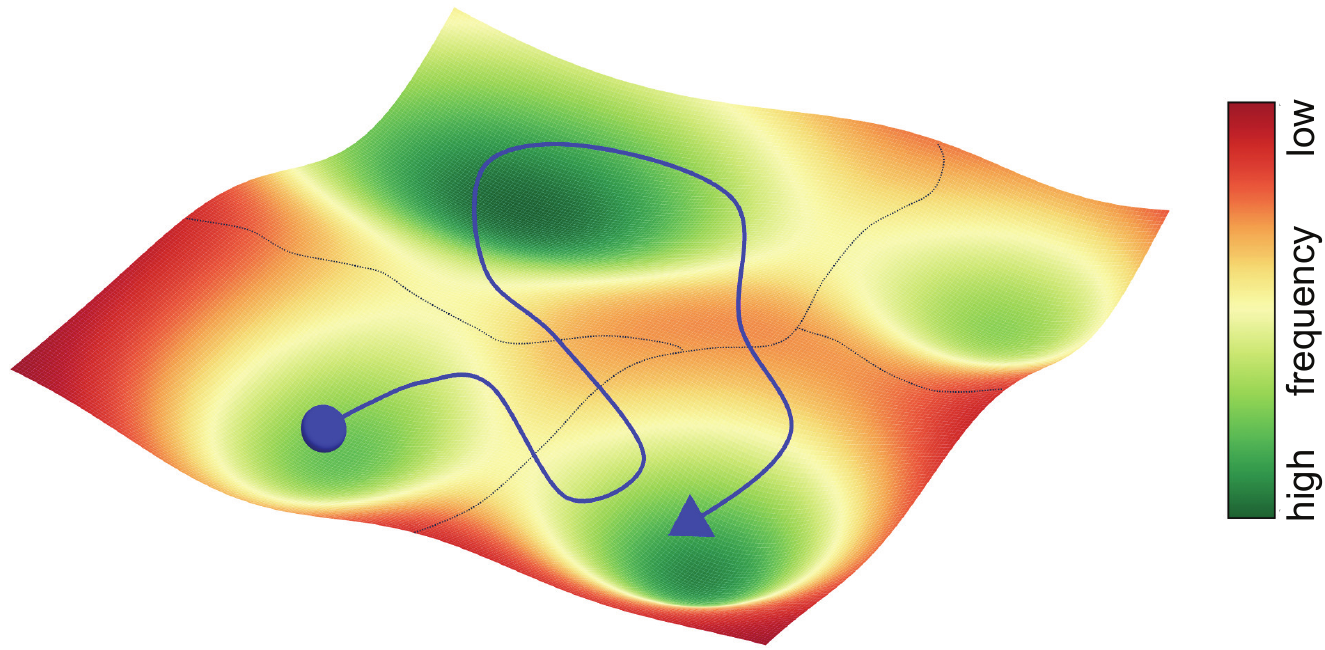}
\caption{Two-dimensional schematic of an energy landscape. Four local minima are present. The dotted lines show the borders between the basin of different local minima. The ``ball'' represents the state of the system, and its position on the plane (ignoring the vertical coordinate) represents the activity pattern.}
\label{fig:2dim-schem}
\end{center}
\end{figure}

Energy landscape is an established concept, catered with computational methods, in protein folding research
\cite{Onuchic1997AnnuRevPhysChem, Wales2003book, Wales2005PhilTransRSocA, Roder2022FrontMolBiosci}. With this dynamic view, we comprehend protein folding as an ensemble of locally stable protein conformations that are stochastically switched between depending on how stable each conformation is and how likely one conformation changes to another. According to the principles of statistical physics, a conformation with a low energy occurs with a high probability, and transitions between conformations occur with a high frequency if the energy barrier separating them is low (see Figs~\ref{fig:1dim-schem} and \ref{fig:2dim-schem} for schematics). The purpose of this review paper is to explain a family of data analysis methods for constructing such energy landscapes and dynamics on them for general multivariate time series data for which we do not know any underlying kinetic, chemical, or other laws.

The type of energy landscape analysis (ELA) that this article deals with, which we simply refer to as ELA in the following text, was initially proposed in \cite{Watanabe2014FrontNeuroinfo, Watanabe2014NatComm} for functional magnetic resonance imaging (fMRI) data recorded from the human brain. It uses the Ising spin system model.
The model fitting part of this ELA is shared with earlier studies, in particular in neuroscience
\cite{Schneidman2006Nature, Shlens2006JNeurosci, Tang2008JNeurosci, YuHuangSinger2008CerebralCortex, Yeh2010Entropy, Mora2011JStatPhys, Watanabe2013NatComm}.
Related to this fact, the ELA has predominantly been used in neuroscience, especially for human fMRI data
\cite{WatanabeRees2017NatComm, Ashourvan2017Neuroimage, Kang2017Neuroimage, Ezaki2018HumanBrainMapping, KangPae2019PlosOne, Ezaki2020CommunBiol, JeongKang2021Neuroimage, KangJeong2021HumanBrainMapping, Regonia2021FrontPsychiatry, Yamashita2021Neuroimage, Yamashita2021SciRep, Kondo2022FrontNeurosci, FanLiHuang2022NeuroimageClinical, LiAnZhou2023FrontNeurosci, Theis2023biorxiv, Watanabe2023Eneuro, Hosaka2024biorxiv, Ishida2024NeuroimageClinical, Khanra2024EurJNeurosci, Kindler2024SchizophreniaBullOpen, Miyata2024PsychiatryClinicalNeurosci, XingGuoLong2024FrontAgingNeurosci}. Other applications in neuroscience include electroencephalogram (EEG) \cite{Klepl2022IeeeJBiomedHealthInfo, Watanabe2021Elife, Gupta2023Sensors}, magnetoencephalography (MEG) \cite{Krzeminski2020NetwNeurosci}, and structural brain network \cite{GuCieslak2018SciRep} data.
However, the ELA does not impose any neuroscientific assumptions on the data and therefore applicable to any multivariate time series data as long as the data are long enough relative to the number of variables. In fact, the ELA
has seen applications to simulated data of cardiac fibrillation \cite{Song2018arxiv}, 
time series of a combination of multimodal observables from patients in rheumatoid arthritis (such as the rheumatoid factor and swollen joint count) \cite{YamamotoSakaguchi2024PlosOne}, Raman spectroscopy data \cite{Yonezawa2024IntJMolSci},
microbiome data \cite{Suzuki2021EcolMonographs, Fujita2022biorxiv, Fujita2023Microbiome, Miyamoto2023EnvironRes}, and mixture
of microbiome data and physiological signals from calves \cite{OkadaInabu2023SciRep}.
ELA crucially takes a system's dynamics point of view in the sense that the method tracks the dynamics of the system's state, which is defined by the values of all variables at each time $t$. Such a system's state is abstracted into the filled ball in Figs~\ref{fig:1dim-schem}(a), \ref{fig:1dim-schem}(b), and \ref{fig:2dim-schem}. As these figures show, ELA aims to capture the given multivariate time series data as the single abstracted system's state that transits among discrete states, similar to a Markov chain. Furthermore, the present ELA is based on the computation and statistical physics theory of the Ising spin system model. There are various other energy landscape analyses in addition to those in protein folding studies (e.g. \cite{KimWang2007PlosComputBiol, Marvel2009PhysRevLett, LvLiLi2015PlosComputBiol, Pradagracia2009PlosComputBiol, Andersson2022Iscience}). In this review paper, we exclusively focus on the ELA based on the Ising model.

Our intention to write this tutorial-prone survey paper is to guide potential users of ELA in various research fields. At the same time, probably owing to open-source code (reviewed in Section~\ref{sec:software}), including some with a graphical interface, one can now easily carry out ELA without knowing details. This situation is beneficial for boosting applications of the method. However, there are recommended practices and pitfalls in ELA, which we will advocate in this paper. Furthermore,
a detailed step-by-step tutorial for non-specialists is expected to help users understand intuitions, physics theory, and logic behind the scene. Therefore, although we published a shorter review in 2017 \cite{Ezaki2017PhilTransRSocA}, here we provide a detailed tutorial, which includes discussion of recommended practices and pitfalls, advantages and disadvantages of ELA, and recent methodological developments, followed by some example applications.

\section{Workflow of the energy landscape analysis}

\subsection{Overview}

\begin{figure}
\begin{center}
\includegraphics[width=12cm]{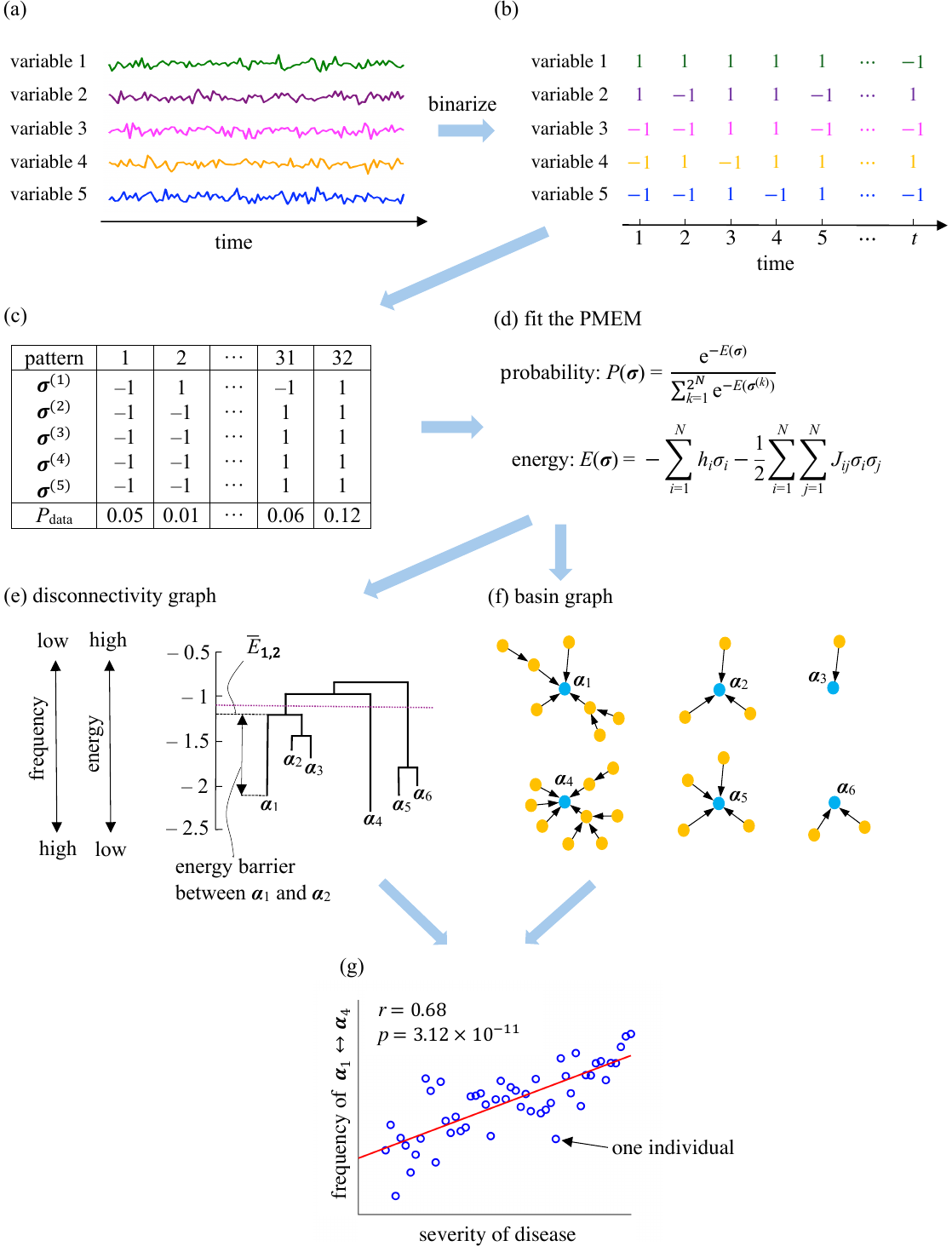}
\caption{Steps of the ELA. (a) The input data is multivariate time series with $N$ variables.
(b) Binarized time series. (c) Fraction of times, $P(\bm{\sigma})$, with which each of the $2^N$ activity patterns, $\bm{\sigma}$, occurs in the binarized data shown in (b). (d) Fitting of the PMEM to the distribution of $\bm{\sigma}$ shown in (c). (e) Disconnectivity graph derived from the estimated PMEM.
It shows relationships between the activity patterns that are local minima.
Each $\bm{\alpha}_i$, $i\in \{1, \ldots, 6 \}$, represents a local minimum. The double-headed arrow indicates the height of the energy barrier between local minima $\bm{\alpha}_1$ and $\bm{\alpha}_2$ when the activity pattern tries to travel from $\bm{\alpha}_1$ to $\bm{\alpha}_2$. (f) Basin graph derived from the estimated PMEM. (g) Downstream analysis, which may use a feature computed from the estimated disconnectivity graph or basin graph. }
\label{fig:pipeline}
\end{center}
\end{figure}

We outline a workflow of ELA in Fig~\ref{fig:pipeline}.
First, the input data are multivaraite time series with $N$ variables in discrete time (Fig~\ref{fig:pipeline}(a)). We denote the time series by $\{\bm{x}(1), \ldots, \bm{x}(t_{\max}) \}$, where $\bm{x}(t)$ (with $t\in \{1, \ldots, t_{\max}\}$) is the $N$-dimensional time series at time $t$, and $t_{\max}$ is the number of observations, or the length of the time series. We denote the $N$-dimensional vector $\bm{x}(t)$ by
\begin{equation}
\bm{x}(t) = (x_1(t), \ldots, x_N(t)),
\end{equation}
where $x_i(t) \in \mathbb{R}$ is the signal of the $i$th variable at time $t$. We usually assume $N$ of the order of 10; we discuss the choice of $N$ in Section~\ref{sub:challenges}.

Second, we binarize each $x_i(t)$, $i\in \{1, \ldots, N\}$, $t\in \{1, \ldots, t_{\max} \}$. See Fig~\ref{fig:pipeline}(b) for an example of binarized multivariate time series. We denote the obtained binarized multivariate time series by $\{ \bm{\sigma}(1), \ldots, \bm{\sigma}(t_{\max}) \}$, where the system's binarized state, which we call the activity pattern, at time $t$ is written as
\begin{equation}
\bm{\sigma}(t) = (\sigma_1(t), \ldots, \sigma_N(t)),
\end{equation}
where $\sigma_i(t) \in \{-1, 1 \}$. Another common convention is $\sigma_i(t) \in \{0, 1 \}$. In either convention, $1$ corresponds to high activity of the $i$th variable at time $t$; the other value (i.e., $-1$ or $0$ depending on the coding scheme) corresponds to low activity. Because the linear transformation $f(x)=(x+1)/2$ maps $\{-1, 1\}$ to $\{0, 1\}$, i.e., $f(-1) = 0$ and $f(1) = 1$, the choice of either convention does not essentially matter for the following analyses. See \cite{Ezaki2017PhilTransRSocA} for more discussion of the two conventions.

Third, we estimate the Ising model, i.e., infer the model parameters from the binarized multivariate time series data $\{\bm{\sigma}(1), \ldots, \bm{\sigma}(t_{\max}) \}$ (Fig~\ref{fig:pipeline}(c) and (d)). The Ising model is also referred to the pairwise maximum entropy model (PMEM), which we use throughout this paper, and the Boltzmann machine. There are various inference algorithms. The estimated Ising model is typically a maximum likelihood estimator for the data under some constraints. In fact, the procedures up to this point, i.e., fitting the Ising model to multivariate time series and other multivariate data, is not a new idea. See
\cite{Yeh2010Entropy, Mora2011JStatPhys, Stein2015PlosComputBiol, Nguyen2017AdvPhys, Carleo2019RevModPhys, Mehta2019PhysRep} %
%
for reviews.

The following steps are unique to the ELA. Fourth, we estimate the so-called disconnectivity graph \cite{Becker1997JChemPhys}, of which a hypothetical example is shown in Fig~\ref{fig:pipeline}(e). The bottom of each valley schematically shown in Figs~\ref{fig:1dim-schem} and \ref{fig:2dim-schem} corresponds to the tip of each branch hanging down vertically in the disconnectivity graph (see Fig~\ref{fig:pipeline}(e)). It is called a local minimum of energy (local minimum for short). The disconnectivity graph represents all the local minima and their relationships in terms of the ease of transition for the estimated PMEM and is a primary product of the ELA. Another main product of the ELA is the basin graph, shown in Fig~\ref{fig:pipeline}(f). The basin graph encodes the width of the valley of each local minimum schematically shown in Fig~\ref{fig:1dim-schem} or its area schematically shown in Fig~\ref{fig:2dim-schem}. It also contains other information.

Fifth, because the disconnectivity graph and basin graph are graphical representations of the energy landscape, we quantify them into numbers or vectors, which we collectively call features of the energy landscape. These features provide insights into the given multivariate time series data or a system generating the data.

Sixth, one often wants to compare the energy landscapes obtained from a group of people or an individual against the energy landscapes obtained from a different group or individual. Such a comparison is made in terms of features of the energy landscape. For example, we examine whether or not larger values of a particular feature from the energy landscape estimated for different human individuals are associated with the severity of a disease (see Fig~\ref{fig:pipeline}(g)). In the following sections, we explain each step in detail.

\subsection{Fitting the PMEM\label{sub:fitting-algorithm}}

We start by binarizing the original time series (see Fig~\ref{fig:pipeline}(a) and (b)). For each $i \in \{1, \ldots, N\}$ and $t \in \{1, \ldots, t_{\max} \}$, we set $\sigma_i(t) = -1$ if $x_i(t) < \theta_i$ and $\sigma_i(t) = 1$ if $x_i(t) \ge \theta_i$, where $\theta_i$ is an arbitrary threshold. A common choice of $\theta_i$ is the 50 percentile of $\{x_i(1), \ldots, x_i(t_{\max}) \}$ such that $\sigma_i(t) = -1$ and $\sigma_i(t) = 1$ with probability $0.5$ each. If the distribution of $x_i(t)$ is not extremely skewed,
one can also set $\theta_i$ to the average of the time series, i.e., $\sum_{t=1}^{t_{\max}} x_i(t) / t_{\max}$, to obtain roughly 50\% of $\sigma_i(t) = -1$ and $\sigma_i(t) = 1$ each.

Because $\bm{\sigma}(t)$ is an $N$-dimensional binary vector, there are $2^N$ possible activity patterns in total, which we enumerate as $\bm{\sigma}^{(1)}$, $\ldots$, $\bm{\sigma}^{(2^N)}$. This is depicted in Fig~\ref{fig:pipeline}(c). For example,
one obtains $\bm{\sigma}^{(1)} = (-1, -1, \ldots, -1)$, $\bm{\sigma}^{(2)} = (1, -1, \ldots, -1)$, and $\bm{\sigma}^{(2^N)} = (1, 1, \ldots, 1)$. We then count the fraction of each activity pattern $\bm{\sigma}$, denoted by $P_{\text{data}}(\bm{\sigma})$, in the given data. Note that $\bm{\sigma}$ is any of $\bm{\sigma}^{(1)}$, $\ldots$, $\bm{\sigma}^{(2^N)}$ and that $P_{\text{data}}(\bm{\sigma})$ is simply the number of times that $\bm{\sigma}$ appears in the data that is divided by $t_{\max}$. Because $P_{\text{data}}(\bm{\sigma})$ is a fraction, we obtain
\begin{equation}
\sum_{k=1}^{2^N} P_{\text{data}}(\bm{\sigma}^{(k)}) = 1.
\end{equation}

The PMEM aims to fit a particular form of the distribution, denoted by $P_{\text{model}}(\bm{\sigma})$ and explained in the following text, to the empirical distribution, $P_{\text{data}}(\bm{\sigma})$. With this methodological choice, we are drastically ignoring the temporal order of the given data. For example,
the estimated PMEM will be the same as that for the original time series even if we re-arrange the original multivariate time series in the reversed order (i.e., $\{ \bm{\sigma}(t_{\max}), \bm{\sigma}(t_{\max}-1), \ldots, \bm{\sigma}(1) \}$) or shuffle the time stamp of $\{ \bm{\sigma}(t) \}$ randomly before the analysis. Therefore, the ELA is in a sense not a method for analyzing time series. It is a method to describe the distribution of the given data. However, ELA still allows us to infer and interpret the \textit{dynamics} of the data; e.g., the system's state (i.e., the ball in Figs~\ref{fig:1dim-schem} and \ref{fig:2dim-schem}) stays near a local minimum $\bm{\alpha}_1$ for some time, then transits to a different local minimum $\bm{\alpha}_2$, and comes back to $\bm{\alpha}_1$. This is because the PMEM implies a particular equilibrium stochastic dynamics. In other words, once a PMEM is estimated, it dictates the probability or rate of the move from any activity pattern $\bm{\sigma}$ to another, which we explain in Section~\ref{sub:random-walk}.

The mean activity of the $i$th variable for the data is given by
\begin{equation}
\langle \sigma_i \rangle_{\text{data}} \equiv \frac{1}{t_{\max}} \sum_{t=1}^{t_{\max}} \sigma_i(t).
\end{equation}
The mean pairwise joint activity for the $i$th and $j$th variables for the data is given by
\begin{equation}
\langle \sigma_i \sigma_j \rangle_{\text{data}} \equiv \frac{1}{t_{\max}} \sum_{t=1}^{t_{\max}} \sigma_i(t) \sigma_j(t).
\end{equation}
The PMEM maximizes the entropy of the distribution of activity patterns, $P(\bm{\sigma})$, under the condition that $\langle \sigma_i \rangle$ and $\langle \sigma_i \sigma_j \rangle$ (with $1 \leq i \leq j \leq N)$ are the same between the estimated Ising model and the given data for all $i, j \in \{1, \ldots, N\}$. As a result of entropy maximization, one obtains (see Fig~\ref{fig:pipeline}(d))
\begin{equation}
P(\bm{\sigma})=\frac{e^{-E(\bm{\sigma})}}{\sum_{k=1}^{2^N} e^{-E(\bm{\sigma}^{(k)})}},
\label{eq:P(V)}
\end{equation}
where $E(\bm{\sigma})$ represents the energy of activity pattern $\bm{\sigma}$ and is given by
\begin{equation}
E(\bm{\sigma})=-\sum_{i=1}^N h_i \sigma_i - \frac{1}{2} \sum_{i=1}^N \sum_{j=1}^N J_{ij} \sigma_i \sigma_j.
\label{eq:E(V)}
\end{equation}
%
%
Equation~\eqref{eq:P(V)} is a Boltzmann distribution (also called Gibbs distribution) in statistical mechanics without the temperature parameter, or equivalently, the temperature normalized to $1$.
Equation~\eqref{eq:P(V)} implies that a high probability of the activity pattern $P(\bm{\sigma})$ corresponds to a low energy $E(\bm{\sigma})$ and vice versa. It should be noted that the energy here is a purely mathematical construct and does not imply any physical energy.
In Eq~\eqref{eq:E(V)}, the fitting parameter $h_i$ represents the tendency that the $i$th variable is active (i.e., $\sigma_i = 1$), and $J_{ij}$ quantifies the sign and strength of the pairwise interaction between the $i$th and $j$th variables. The PMEM assumes that $J_{ij} = J_{ji}$. Furthermore, the $J_{ij} \sigma_i \sigma_j$ when $i=j$ is reduced to a constant $J_{ii} \sigma_i^2 = J_{ii}$ because $\sigma_i^2$ with $\sigma_i \in \{ -1, 1 \}$ is always equal to $1$. Because $P(\bm{\sigma})$ in Eq~\eqref{eq:P(V)} does not change by the addition of a constant term to $E(\bm{\sigma})$, one can simplify Eq~\eqref{eq:E(V)} to
\begin{equation}
E(\bm{\sigma})=-\sum_{i=1}^N h_i \sigma_i - \sum_{i=1}^N \sum_{j=1}^{i-1} J_{ij} \sigma_i \sigma_j.
\label{eq:E(V)-simplified}
\end{equation}

We denote the expected single-variable activity and pairwise activity from the PMEM by $\langle \sigma_i \rangle_{\rm model}$ and $\langle \sigma_i \sigma_j \rangle_{\rm model}$, respectively. They are given by
\begin{equation}
\langle \sigma_i \rangle_{\rm model}= \sum_{k=1}^{2^N} \sigma_i^{(k)} P(\bm{\sigma}^{(k)})
\end{equation}
and
\begin{equation}
\langle \sigma_i \sigma_j \rangle_{\rm model} = \sum_{k=1}^{2^N} \sigma_i^{(k)} \sigma_j^{(k)} P(\bm{\sigma}^{(k)}),
\end{equation}
where $\sigma^{(k)}_i \in \{-1, 1\}$ is the binarized activity of the $i$th variable in the $k$th activity pattern $\bm{\sigma}^{(k)}$.

A popular and conceptually easy method to infer $h_i$ and $J_{ij}$ is to iteratively adjusting 
them using a gradient ascent algorithm; it is an ascent algorithm because it gradually increases the likelihood of the data, $\mathcal{L} = \prod_{t=1}^{t_{\max}} P(\bm{\sigma}(t))$ along the ascent direction. The iteration scheme is given by
\begin{equation}
h_i^{\text{new}} =h_i^{\text{old}} + \frac{\epsilon}{t_{\max}} \frac{\partial \mathcal{L}} {\partial h_i}
= h_i^{\text{old}} + \epsilon (\langle \sigma_i \rangle_{\text{data}} - \langle \sigma_i 
\rangle_{\rm model})
\label{eq:h_i-update}
\end{equation}
and
\begin{equation}
J_{ij}^{\text{new}}=J_{ij}^{\text{old}} + \frac{\epsilon}{t_{\max}} \frac{\partial \mathcal{L}} {\partial J_{ij}}
= J_{ij}^{\text{old}} + \epsilon (\langle \sigma_i \sigma_j \rangle_{\text{data}} - \langle \sigma_i \sigma_j \rangle_{\rm model}),
\label{eq:J_{ij}-update}
\end{equation}
where superscripts new and old represent the values after and before a single updating step, respectively, and $\epsilon$ is the learning rate. The factor $1/t_{\max}$ on the right-hand sides of Eqs~\eqref{eq:h_i-update} and \eqref{eq:J_{ij}-update} is for normalization. 
A slightly different iteration scheme called the iterative scaling algorithm \cite{Darroch1972AnnMathStat}, in which the right-hand side of Eqs~\eqref{eq:h_i-update} and \eqref{eq:J_{ij}-update} is replaced by $\epsilon \sgn(\langle \sigma_i\rangle)\log (\langle \sigma_i\rangle_{\text{data}}/\langle \sigma_i\rangle_{\rm{model}})$ and $\epsilon \sgn(\langle \sigma_i\sigma_j\rangle)\log (\langle \sigma_i\sigma_j\rangle_{\text{data}}/\langle \sigma_i\sigma_j\rangle_{\rm{model}})$, respectively, is also common \cite{Schneidman2006Nature, Yeh2010Entropy, Watanabe2013NatComm}.
With either iterative algorithm, the convergence (i.e., $h_i^{\text{new}} =h_i^{\text{old}}$ and  
$J_{ij}^{\text{new}}=J_{ij}^{\text{old}}$ for all $i, j$) implies that 
$\langle \sigma_i \rangle_{\text{model}} = \langle \sigma_i  \rangle_{\rm data}$ and
$\langle \sigma_i \sigma_j \rangle_{\text{model}} = \langle \sigma_i \sigma_j \rangle_{\rm data}$
for all $i, j$.

One can verify that Eq~\eqref{eq:P(V)} is concave in terms of each $h_i$ and $J_{ij}$ by checking that the Hessian of $\log \mathcal{L}$ is a type of sign-flipped covariance matrix, which is negative semi-definite. The concavity guarantees that the gradient ascent scheme always yields the unique maximum likelihood estimator.
However, a single updating step requires reference to $P(\bm{\sigma})$ for all the $2^N$ activity patterns for calculating $\langle \sigma_i \rangle_{\rm{model}}$ and $\langle \sigma_i\sigma_j \rangle_{\rm{model}}$. Because
$P(\bm{\sigma})$ changes in every iteration step, the gradient ascent method is computationally costly for large $N$. Our experience is that, without dedicated hardware, the gradient ascent method would be slow even with $N=15$ or 20. To circumvent this problem, there are various approximate algorithms (e.g., pseudo-likelihood maximization) for estimating the PMEM \cite{Besag1975JRStatSocserD, Roudi2009FrontComputNeurosci, Stein2015PlosComputBiol, Ezaki2017PhilTransRSocA, Nguyen2017AdvPhys}.

\subsection{Accuracy of fit\label{sub:accuracy-fit}}

If the PMEM poorly fits to the data, ELA is probably not a good choice. Therefore, while many papers skip measuring an accuracy of fit, it is important to measure it.

A least stringent test, which is often used, is to examine the correlation between $\langle \sigma_i \rangle_{\text{model}}$ and $\langle \sigma_i \rangle_{\text{data}}$ across the $i$ values and that between 
$\langle \sigma_i \sigma_j \rangle_{\text{model}}$ and $\langle \sigma_i \sigma_j \rangle_{\text{data}}$ across $(i, j)$ pairs.
A good fit of the PMEM to the data should yield
$\langle \sigma_i \rangle_{\text{model}} \approx \langle \sigma_i \rangle_{\text{data}}$ and
$\langle \sigma_i \sigma_j \rangle_{\text{model}} \approx \langle \sigma_i \sigma_j \rangle_{\text{data}}$, where $\approx$ represents ``approximately equal to''.
However, these relationships are not enough for the following reason. If we run a gradient ascent algorithm to fit the PMEM (Section~\ref{sub:fitting-algorithm}), it will converge due to the concavity of the problem for any data while convergence may take a long time if $N$ is relatively large. Once the algorithm sufficiently converges, one obtains
$\langle \sigma_i \rangle_{\text{model}} \approx \langle \sigma_i \rangle_{\text{data}}$ and
$\langle \sigma_i \sigma_j \rangle_{\text{model}} \approx \langle \sigma_i \sigma_j \rangle_{\text{data}}$ even if the data are a poor fit to the PMEM. 

Therefore, one often plots $P(\bm{\sigma})$ against $P_{\text{data}}(\bm{\sigma})$ over all the $2^N$ activity patterns $\bm{\sigma}$. If $P(\bm{\sigma})$ and $P_{\text{data}}(\bm{\sigma})$ are highly correlated, the accuracy of the fit is considered to be high
\cite{Schneidman2006Nature, Shlens2006JNeurosci, Tang2008JNeurosci, YuHuangSinger2008CerebralCortex, Mora2011JStatPhys, Watanabe2013NatComm, Watanabe2014NatComm, Watanabe2014FrontNeuroinfo, Ezaki2018HumanBrainMapping}.
%

Another commonly used accuracy measure~\cite{Shlens2006JNeurosci, Yeh2010Entropy, Ezaki2017PhilTransRSocA} is
\begin{equation}
r_D=\frac{D_1-D_2}{D_1},
\end{equation}
where
\begin{equation}
D_{\ell}=\sum_{k=1}^{2^N}P_{\text{data}}(\bm{\sigma}^{(k)}) \log_2 \frac{P_{\text{data}}(\bm{\sigma}^{(k)})}{P_{\ell}(\bm{\sigma}^{(k)})}
\label{eq:D_ell-def}
\end{equation}
is the Kullback-Leibler divergence between the probability distribution of the activity pattern in the $\ell$th-order $(\ell=1, 2)$ maximum entropy model, $P_{\ell}(\bm{\sigma})$, and the distribution of the activity pattern in the original data, $P_{\text{data}}(\bm{\sigma})$. Distribution $P_2(\bm{\sigma})$ is equivalent to $P(\bm{\sigma})$ given by Eqs~\eqref{eq:P(V)} and \eqref{eq:E(V)}. It is called second-order because the highest-order function in terms of $\sigma_i$ appearing in the model is of second order (i.e., $J_{ij} \sigma_i \sigma_j$).
The first-order, or independent, maximum entropy model (i.e., $\ell=1$) is given by Eq~\eqref{eq:P(V)} but without interaction terms. In other words, we force $J_{ij}=0$ $\forall i,j$ in Eq~\eqref{eq:E(V)} and set each $h_i$ so that $\langle \sigma_i \rangle_{\text{model}} = \langle \sigma_i \rangle_{\text{data}}$ is satisfied. Quantity $r_D$ ranges between $0$ and $1$ for the maximum likelihood estimator of the PMEM. We obtain $r_D=1$ when the PMEM perfectly fits the distribution of the activity pattern for the given data. In contrast, $r_D=0$ when the PMEM does not fit the data any better than the first-order model.
For the maximum likelihood estimator, $r_D$ is equal to another often employed accuracy measure given by $(S_1 - S_2)/(S_1 - S_{\text{data}})$, where
$S_{\ell} = - \sum_{k=1}^{2^N} P_{\ell}(\bm{\sigma}^{(k)}) \log P_{\ell}(\bm{\sigma}^{(k)})$ is the Shannon entropy of distribution $P_{\ell}$ 
\cite{Tang2008JNeurosci, Yeh2010Entropy}.

The accuracy of fit would be small when $N$ is large or $t_{\max}$ is small. In fact, the ELA is a data-hungry method. Therefore, it is important to check and report the value of $r_D$ or a different accuracy of fit measure. We recommend the value of $r_D$ above $0.85$, or ideally $0.9$. This is because, although these guideline values do not have any theoretical ground, an influential early study of the PMEM applied to neural spiking data realized $r_D \approx 0.9$ \cite{Schneidman2006Nature}. 

\subsection{Local minima of energy and the network of activity patterns}

Once we have estimated the PMEM, the next task is to construct and draw the so-called disconnectivity graph. To understand the meaning and construction of the disconnectivity graph, we need to introduce two concepts: local minimum of energy and network of activity patterns.

\begin{figure}[t]
\begin{center}
\includegraphics[width=8cm]{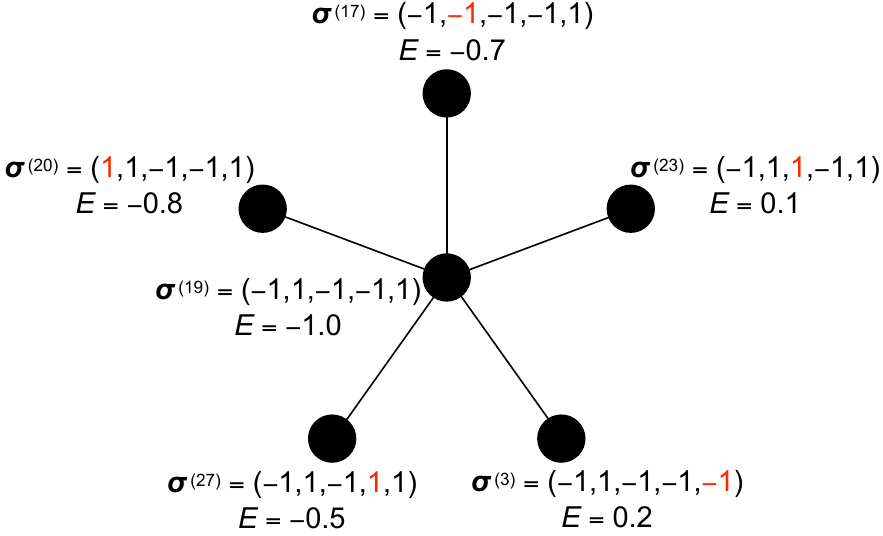}
\caption{A schematic of the local minimum of energy. A system of $N=5$ variables is assumed. Activity pattern $\bm{\sigma}^{(19)}$ is a local minimum because it has a lower energy than all its neighbors. The neighbors are the activity patterns with Hamming distance 1 from $\bm{\sigma}^{(19)}$. The numbering scheme for the activity patterns is consistent with the one used in Fig~\ref{fig:pipeline}(c).}
\label{fig:neighborhood}
\end{center}
\end{figure}

Intuitively, a local minimum of energy is the bottom of a basin of the energy landscape shown in Figs~\ref{fig:1dim-schem} and \ref{fig:2dim-schem}. However, the $2^N$ activity patterns do not lie in a one- or two-dimensional space as depicted in these figures. To precisely define the local minimum, let us consider Fig~\ref{fig:neighborhood}, in which there are $N=5$ variables and an activity pattern $\bm{\sigma}^{(19)} = (-1, 1, -1, -1, 1)$ is neighbored by five other activity patterns. Each of these neighboring activity patterns is different from $\bm{\sigma}^{(19)}$ in the sign of just one $\sigma_i$. In other words, these five activity patterns are in Hamming distance $1$ from $\bm{\sigma}^{(19)}$. There are five activity patterns neighboring $\bm{\sigma}^{(19)}$ because one can flip any one entry of $\bm{\sigma}^{(19)}$ to obtain a neighboring activity pattern.
Fig~\ref{fig:neighborhood} shows that $\bm{\sigma}^{(19)}$ has a lower energy than all the five neighbors. Therefore, $\bm{\sigma}^{(19)}$ appears more frequently than each of its neighboring activity patterns in the data. When this is the case, we say that $\bm{\sigma}^{(19)}$ is a local minimum. If we allow the activity pattern to vary over time, the system's state is more likely to transit from a neighbor of $\bm{\sigma}^{(19)}$, say $\bm{\sigma}^{(17)}$, to $\bm{\sigma}^{(19)}$ than vice versa because $\bm{\sigma}^{(19)}$ is more frequent than $\bm{\sigma}^{(17)}$. In this sense, $\bm{\sigma}^{(19)}$ is locally attractive.

However, such dynamics are by definition stochastic. Therefore, $\bm{\sigma}^{(19)}$ being a local minimum implies that the system's state arriving at $\bm{\sigma}^{(19)}$ will still escape it albeit at a lower rate. In empirical data, we commonly see transitions out of local minima.

\begin{figure}[t]
\begin{center}
\includegraphics[width=14cm]{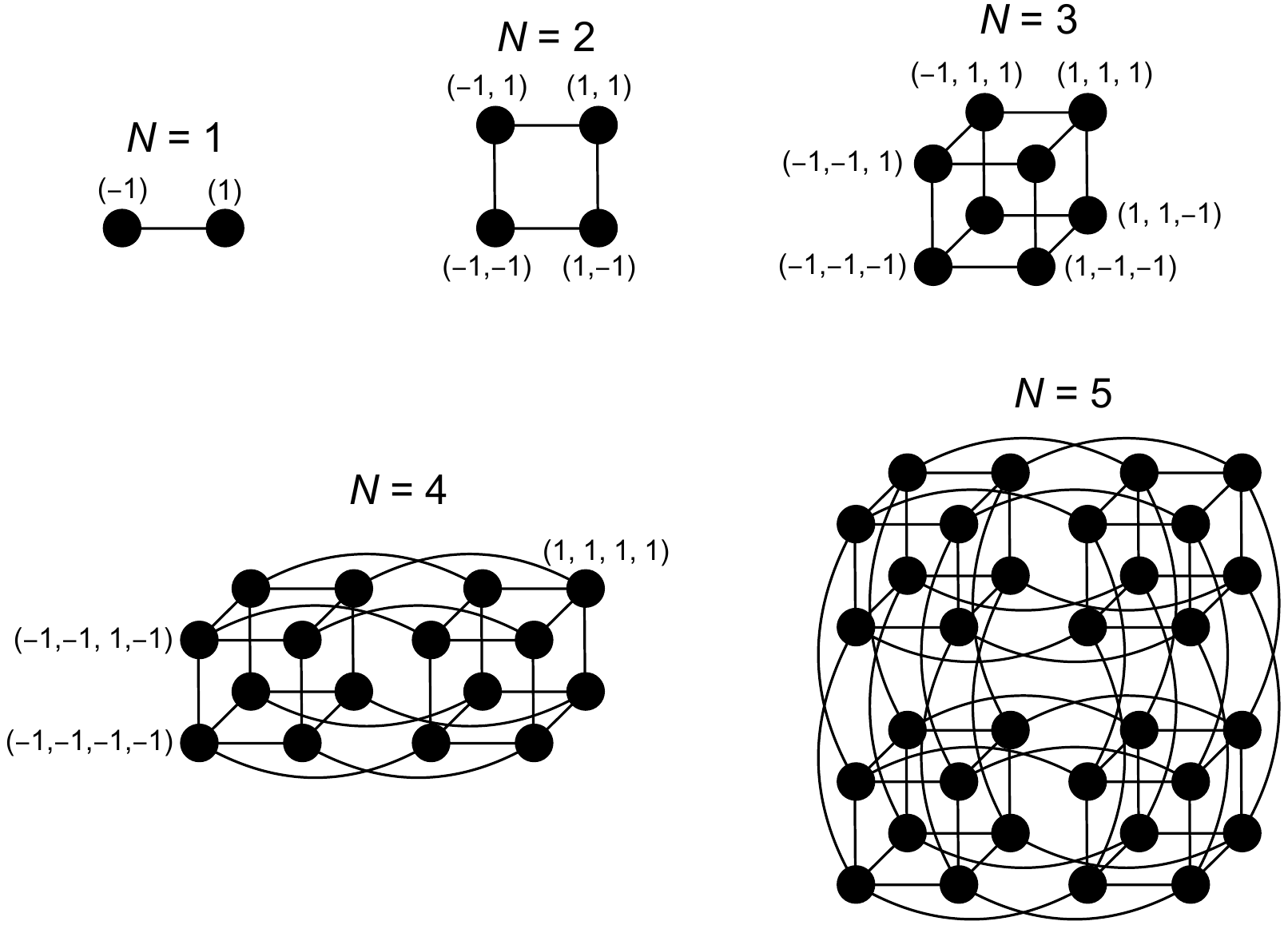}
\caption{Networks of activity patterns for $N \in \{1, \ldots, 5 \}$. They are hypercubes having $2^N$ nodes. Each node represents an activity pattern and has $N$ neighbors.}
\label{fig:hypercubes}
\end{center}
\end{figure}

In fact, any activity pattern has $N$ neighbors no matter whether or not it is a local minimum. The $2^N$ activity patterns form a ``network of activity patterns'' shown in Fig~\ref{fig:hypercubes}. This network is called the hypercube. One can verify that each node of this network, corresponding to an activity pattern, has exactly $N$ neighbors. 
Using the terminology of network science (or graph theory), there are $2^N$ nodes, and the degree of each node (i.e., the number of edges the node has) is equal to $N$.
Furthermore, in Fig~\ref{fig:hypercubes}, each node has ten nodes in distance (i.e., minimum number of hops) 2, ten nodes in distance 3, five nodes in distance 4, and one node in distance 5. The single node in distance 5 is the one in which all the $N$ values of $\sigma_i$ are flipped. For example, the node in distance 5 from $\bm{\sigma}^{(19)}=(-1, 1, -1, -1, 1)$ is $\bm{\sigma}^{(14)} = (1, -1, 1, 1, -1)$. In general, there are $\binom{N}{d}$ nodes in distance $d$ from any node, where $\binom{}{}$ represents the binomial coefficient. The network of activity patterns represents admissible routes of the system's state, $\bm{\sigma}(t)$, when $\bm{\sigma}(t)$ is allowed to change in one entry, $i$, at one time. 

In empirical data, $\bm{\sigma}(t)$ and $\bm{\sigma}(t+1)$ are often different in more than one $i$ unless the time resolution of the observation is high. For example, in Fig~\ref{fig:pipeline}(b), $\bm{\sigma}(1) = (1, 1, -1, -1, -1)$ and $\bm{\sigma}(2) = (1, -1, -1, 1, -1)$ are different in two variables, i.e., $i=2$ and $i=4$.
If $\bm{\sigma}(t)$ and $\bm{\sigma}(t+1)$ are indeed different in more than one $i$ for some $t$, the trajectory is not a discrete-time walk on the hypercube because the walker can move from a node to its non-neighboring node in one time step.
However, even in this situation, the assumption that only a single entry of $\bm{\sigma}(t)$ can flip at a time is still plausible by considering that we are observing a continuous-time walk in discrete time. This is the case if synchronous threshold crossing, which would simultaneously flip multiple $\sigma_i(t)$, is absent for the resolution of discrete times $t=1$, $2$, $\ldots$ at which one observes the continuous-time walk.

\begin{figure}[t]
\begin{center}
\includegraphics[width=1.0\textwidth]{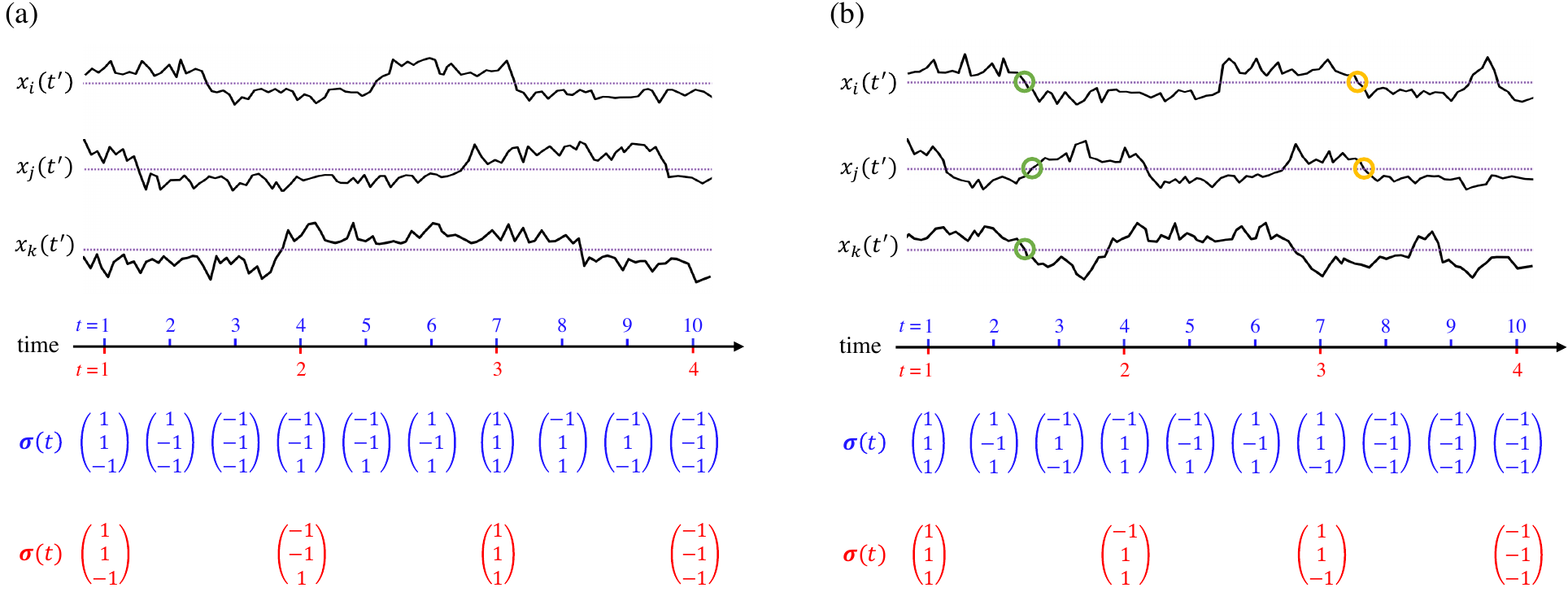}
\caption{Time series of activity pattern as a discrete-time observation of a continuous-time walk. (a) Absence and (b) presence of synchronous threshold crossing. The horizontal dotted lines represent the threshold for binarization, $\theta_i$. We show the case $t=t'$ and $t=3t' - 2$ in blue and red, respectively. In (b), the green circles show an approximately simultaneous threshold crossing of $x_i(t')$, $x_j(t')$, and $x_k(t')$. The yellow circles show an approximately simultaneous threshold crossing of $x_i(t')$ and $x_j(t')$.}
\label{fig:continuous-time-walk}
\end{center}
\end{figure}

To discuss the interpretation of $\bm{\sigma}(t)$ as a discrete-time observation of a continuous-time walk,
we postulate an underlying $N$-dimensional dynamics in continuous time,  $\bm{x}(t')$, $t' \in \mathbb{R}$, and consider that we observe it in discrete times $t \in \{1, 2, \ldots \}$, obtaining $\bm{\sigma}(t)$. Discrete time $t$ does not have to be the same as $t'$.
In Fig~\ref{fig:continuous-time-walk}, the $t$ shown in blue is identical with $t'$. In contrast, the $t$ shown in red is sampled three times less frequently than the $t$ shown in blue.
Recall that $\sigma_i(t)$ flips if and only if $x_i(t)$ crosses the threshold $\theta_i$, no matter whether the time is continuous or discrete. If different $x_i(t)$'s simultaneously cross their thresholds, then multiple $\sigma_i(t)$'s flip at the same time. In this case, $\{ \bm{\sigma}(1), \bm{\sigma}(2), \ldots \}$ is not a nearest-neighbor walk on the hypercube. On the contrary, if such a synchronous threshold crossing is absent and the time resolution for sampling $\bm{\sigma}(t)$ is high enough, then $\bm{\sigma}(t)$ flips at most one bit at a time. In this case, $\{\bm{\sigma}(1), \bm{\sigma}(2), \ldots \}$ is a nearest-neighbor walk. In other words, the walker stays at one node of the hypercube for some discrete times, then moves to one of its neighbors, and so on, producing a discrete-time walk on the hypercube, without jumps from a node to a non-neighboring node. For example, the walk shown in blue in Fig~\ref{fig:continuous-time-walk}(a) is a nearest-neighbor walk. The figure indicates that $x_j(t')$ changes from positive to negative first, followed by the same change in $x_i(t')$. Then, $x_k(t')$ changes from negative to positive. Then, $x_i(t')$, $x_k(t')$, and $x_j(t')$ turn from positive to negative in this order. The discrete-time walk $\bm{\sigma}(t)$ shown in blue captures all these events including their chronology. In contrast, the walk shown in red in Fig~\ref{fig:continuous-time-walk}(a) is not a nearest-neighbor walk because the time resolution for observing $\bm{\sigma}(t)$ is too low. For example, both $\sigma_1(t)$ and $\sigma_2(t)$ change between $t=1$ and $t=2$.

In the PMEM fitted to empirical data, we often find moves from a node on the hypercube to its non-nearest-neighbor, as shown in  Fig~\ref{fig:pipeline}(b) and as $\bm{\sigma}(t)$ shown in red in Fig~\ref{fig:continuous-time-walk}(a). We justify this situation by regarding such $\bm{\sigma}(t)$ as an observation of originally continuously varying $\bm{x}(t')$ at a relatively low time resolution, if there is no approximately simultaneous threshold crossing event.

Fig~\ref{fig:continuous-time-walk}(b) shows a case with approximately simultaneous threshold crossing events in the time series $\{ \bm{x}(t') ; t' \in \mathbb{R} \}$, marked by the green and yellow circles. In this case, even if the time resolution for observing $\bm{\sigma}(t)$ is high, $\{\bm{\sigma}(1), \bm{\sigma}(2), \ldots \}$ is not a nearest-neighbor walk. For example, the walker shown in blue
in Fig~\ref{fig:continuous-time-walk}(b) moves to a non-neighboring node between $t=2$ and $t=3$ and between $t=7$ and $t=8$, which are two intervals in which approximately simultaneous threshold crossing occurs. In fact, a move to a non-neighboring node also occurs between $t=6$ and $t=7$, which contains a less stringent simultaneous threshold crossing of $x_j(t')$ and $x_k(t')$. We also note that the same $\bm{\sigma}(t)$, shown in blue, fails to capture the two threshold crossings of $x_k(t')$ between $t'= t = 8$ and $t' = t = 9$ and two threshold crossings of $x_i(t')$ between $t' = t = 9$ and $t' = t = 10$. In this situation, it is difficult to justify the use of the PMEM and hence ELA.

Approximately simultaneous threshold crossing is a relative concept. 
If we observe $\bm{x}(t')$, $t' \in \mathbb{R}$, at large regular intervals to produce $\bm{\sigma}(t)$, then $x_i(t')$ with multiple $i$'s would cross their thresholds, perhaps multiple times for some $i$'s, between adjacent discrete times (i.e., between $t$ and $t+1$). This is a case of simultaneous threshold crossing.
In the extreme case, if the observation occurs with excessively long time intervals, then $\bm{x}(t)$ and $\bm{x}(t+1)$ would be completely uncorrelated, and so are $\bm{\sigma}(t)$ and $\bm{\sigma}(t+1)$. Then, the trajectory $\{ \bm{\sigma}(1), \bm{\sigma}(2), \ldots \}$ will be composed of independent samples drawn from the distribution $P(\bm{\sigma})$, completely ignoring the hypercube structure.
Given this discussion, we could try to justify the use of the PMEM and ELA even in the case of $\bm{\sigma}(t)$ shown in blue and red in Fig~\ref{fig:continuous-time-walk}(b). If we use a higher sampling rate, such as $t = 100 t'$, such that $t \in \{ 1, 2, \ldots \}$ corresponds to
$t' \in \{0.01, 0.02, \ldots \}$, then $\{ \bm{\sigma}(1), \bm{\sigma}(2), \ldots \}$ will be a nearest-neighbor walk unless there are perfectly simultaneous threshold crossing events in $\bm{x}(t')$, $t' \in \mathbb{R}$. However, excessively synchronous threshold crossing of $x_i(t')$'s would violate the detailed balance condition underlying the random walk assumption (Section~\ref{sub:random-walk}). Therefore, we do not recommend that users rely on random-walk views when strongly synchronous dynamics among the variables is known or expected.

The network of activity patterns should not be confused with networks that variables of the multivariate time series form. One can form a network of the $N$ variables by, for example, naively laying an edge $(i, j)$ if and only if the Pearson correlation coefficient between the two time series $\{x_i(1), \ldots, x_i(t_{\max}) \}$ and $\{x_j(1), \ldots, x_j(t_{\max}) \}$ is sufficiently large \cite{Masuda2023arxiv-corr}. In such a correlation network, a node is a variable of the multivariate time series, and there are $N$ nodes. There may also be a structural network generating $\{ \bm{x}(t) \}$ (e.g., a neuronal network with neurons as nodes, or a social network with human individuals as nodes). In contrast,
in the network of activity patterns, a node is an activity pattern, there are $2^N$ nodes, and the network structure is always the hypercube.

\subsection{Disconnectivity graph}\label{sub:disconnectivity-graph}

\begin{figure}[t]
\begin{center}
\includegraphics[width=14cm]{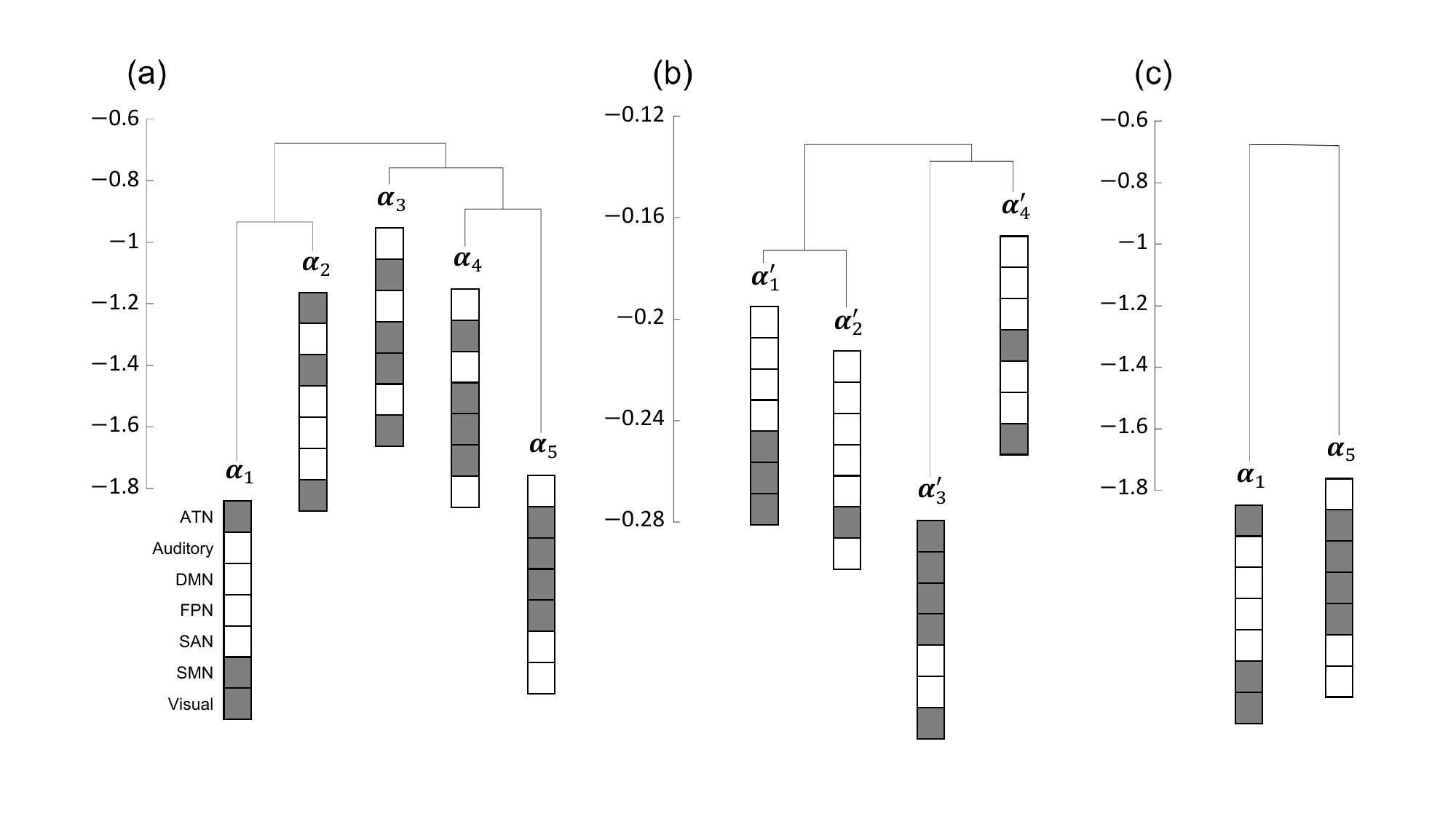}
\caption{Disconnectivity graphs. (a) Disconnectivity graph from fMRI data obtained from the brain of healthy human participants. We used the same data as the one used for Fig A1(a) in \cite{Khanra2024EurJNeurosci}. It is a concatenation of the first four fMRI recording sessions (resulting in $t_{\max} = 818 \times 4 = 3,272$) from a single participant in the Midnight Scan Club data set \cite{Gordon2017Neuron}. We used the first participant (i.e., MSC01). We used the so-called whole-brain network with $N=7$ nodes; the $N$ variables are 
the attention network (ATN), auditory network, default mode network (DMN),
fronto-parietal network (FPN), salience network (SAN),
somatosensory and motor network (SMN), and visual network \cite{WatanabeRees2017NatComm}.
See \cite{Khanra2024EurJNeurosci} for more details and the procedures for preprocessing the fMRI data.
The activity pattern at each local minimum is indicated below the local minimum; a gray square represents $\sigma_i = -1$, and a white square represents $\sigma_i = 1$.
The accuracy of fit $r_D = 0.915$.
%
%
(b) Disconnectivity graph for randomly generated data with the same $N$ and $t_{\max}$ as those used in (a).
(c) Disconnectivity graph obtained via pruning of (a). Local minima $\bm{\alpha}_1$ and $\bm{\alpha}_5$ in the disconnectivity graph shown in (a) have survived the significance test, whereas the other three local minima have not.
}
\label{fig:disconnectivity-graph-MSC}
\end{center}
\end{figure}

The disconnectivity graph is a first-hand and main summary of the energy landscape. The disconnectivity graph is a tool from chemical physics and represents the relationship between different local minima of energy \cite{Becker1997JChemPhys, Wales1998Nature}. For expository purposes, we show a hypothetical disconnectivity graph in Fig~\ref{fig:pipeline}(e); see Fig~\ref{fig:disconnectivity-graph-MSC}(a) for the disconnectivity graph for a real fMRI data set from the human brain.

\begin{figure}[t]
\begin{center}
\includegraphics[width=14cm]{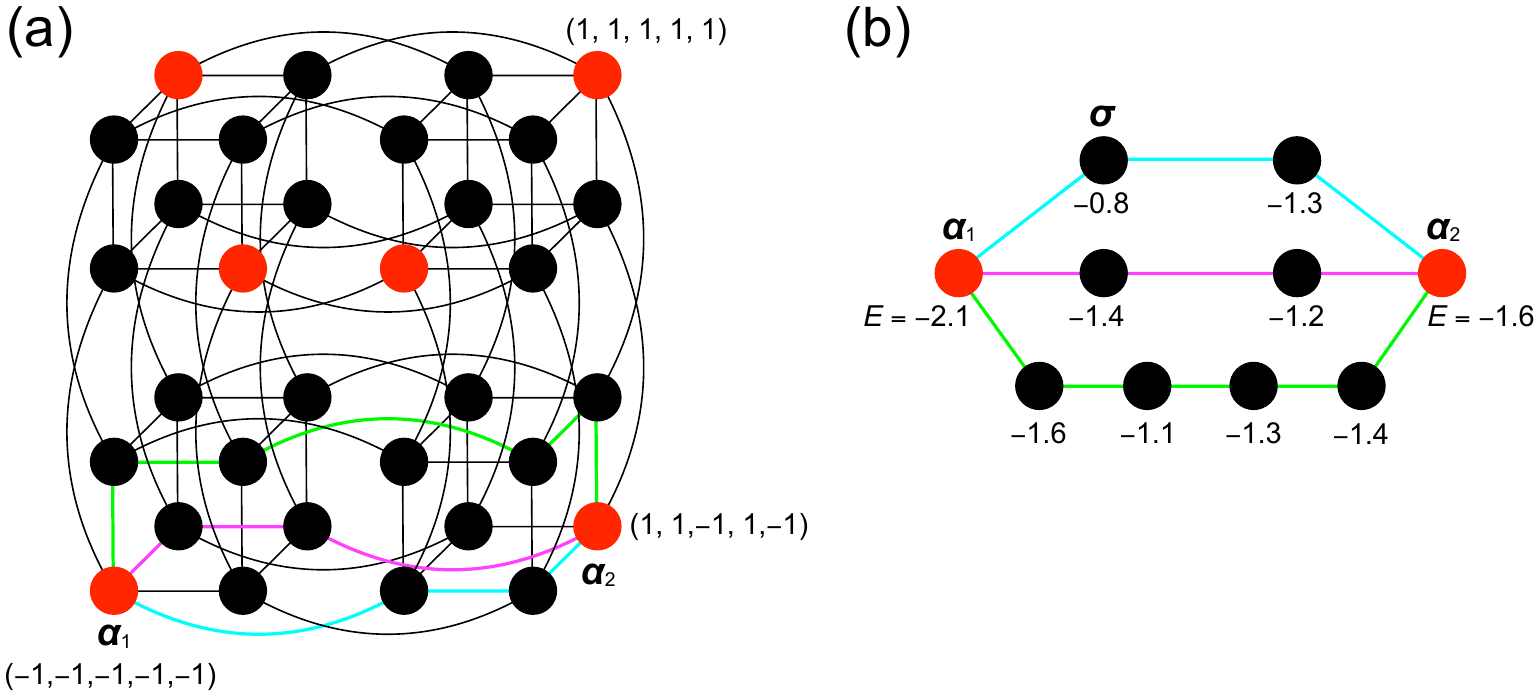}
\caption{Local minima and paths between them. (a) The network of activity patterns in the case of $N=5$ variables. The red circles represent local minima. (b) Three paths between local minima $\bm{\alpha}_1$ and $\bm{\alpha}_2$ in the network of activity patterns shown in (a).}
\label{fig:hypercube-with-paths}
\end{center}
\end{figure}

There may be multiple local minima in the network of activity patterns. In the hypothetical network of activity patterns for $N=5$ variables shown in Fig~\ref{fig:hypercube-with-paths}(a), we indicate the local minima by red circles. To construct the disconnectivity graph, we need to determine the energy barrier separating each pair of local minima. This task would be easy if the energy landscape were one-dimensional, as shown in Fig~\ref{fig:1dim-schem}(a) and (b) because there is only one shortest (i.e., non-redundant) path connecting the two local minima. In fact, this task is more involved because the actual energy landscape of the PMEM is defined on the hypercube. Let us consider possible paths between each pair of local minima. Partly because such a path does not have to be a shortest path (whose length is equal to the Hamming distance between the two local minima), there are many paths unless $N$ is small. The non-black lines in Fig~\ref{fig:hypercube-with-paths}(a) show three among many paths connecting local minima $\bm{\alpha}_1$ and $\bm{\alpha}_2$. Because the Hamming distance between $\bm{\alpha}_1$ and $\bm{\alpha}_2$ is three, any path has at least length 3. For example, the path shown in green has length 5.

Then, to determine the ease with which the system's state can travel from $\bm{\alpha}_1$ to $\bm{\alpha}_2$ and vice versa, let us tabulate all the paths between them and the energy value of each node on the path. 
One can omit walks visiting the same node more than once because such redundant walks do not make the travels between $\bm{\alpha}_1$ and $\bm{\alpha}_2$ easier than the corresponding shortest path.
Fig~\ref{fig:hypercube-with-paths}(b) shows the three paths shown in color in Fig~\ref{fig:hypercube-with-paths}(a) together with the energy on the nodes. The energy value of each node on the path dictates how much height of the hill the system's state needs to climb to go from one end to the other. This is because, by definition, a high-energy activity pattern is a rare activity pattern in the data or the estimated PMEM. For example, in the path shown in skyblue, the highest hill to be overcome is activity pattern $\bm{\sigma}$, which has energy $E=-0.8$. The magenta path makes the travel easier with the largest energy of $E=-1.2$. Although the green path is longer than the skyblue path, the green path has the largest energy of $E=-1.1$, which is smaller than that for the skyblue path. As this example shows, longer paths may be easier to go through
than shorter paths. If all the other paths between $\bm{\alpha}_1$ and $\bm{\alpha}_2$ have an activity pattern with $E > -1.2$, we conclude that the magenta path is the easiest path between $\bm{\alpha}_1$ and $\bm{\alpha}_2$ and record its peak height, denoted by $\overline{E}_{1,2}$ ($= -1.2$ in the current example). Computationally, we do not need to actually enumerate all the possible paths because there are efficient algorithms to realize the same computation \cite{Zhou2011PhysRevLett, Watanabe2014NatComm, Ezaki2017PhilTransRSocA, Khanra2024EurJNeurosci}.

How easy it is for the system's state to travel between $\bm{\alpha}_i$ and $\bm{\alpha}_j$ depends on the energy of $\bm{\alpha}_i$ and $\bm{\alpha}_j$ as well as on $\overline{E}_{i,j}$. If $E(\bm{\alpha}_i)$ is much smaller than $\overline{E}_{i,j}$, then it is difficult for a trajectory starting at $\bm{\alpha}_i$ to reach $\overline{E}_{i, j}$ even if $\overline{E}_{i, j}$ is small. We define the energy barrier between $\bm{\alpha}_i$ and $\bm{\alpha}_j$ from the viewpoint of $\bm{\alpha}_i$ by $\overline{E}_{i, j} - E(\bm{\alpha}_i)$. The energy barrier between the same pair of local minima from the viewpoint of $\bm{\alpha}_j$, which is equal to $\overline{E}_{i, j} - E(\bm{\alpha}_j)$, is different from that from the viewpoint of $\bm{\alpha}_i$ in general. For example, Fig~\ref{fig:hypercube-with-paths}(b) indicates that
$\overline{E}_{1, 2} - E(\bm{\alpha}_1) = -1.2 - (-2.1) = 0.9$ and 
$\overline{E}_{1, 2} - E(\bm{\alpha}_2) = -1.2 - (-1.6) = 0.4$.

We carry out the same computation between each pair of local minima to obtain all $\overline{E}_{i, j}$. Once this is done, one can visualize all the local minima and energy barrier between each pair of them in a dendrogram, which is the disconnectivity graph, as shown in Figs~\ref{fig:pipeline}(e) and \ref{fig:disconnectivity-graph-MSC}(a). 

The vertical axis of the disconnectivity graph represents the energy. A leaf, which vertically hangs down in the upside-down tree, represents a local minimum. The vertical position of the tip of the leaf represents the energy of the local minimum. Observe in 
Fig~\ref{fig:pipeline}(e) that $E(\bm{\alpha}_1) = -2.1$ and $E(\bm{\alpha}_2) = -1.6$, which are consistent with Fig~\ref{fig:hypercube-with-paths}(b). We also observe that $\bm{\alpha}_4$ is the global minimum activity pattern in terms of the energy. Therefore, $\bm{\alpha}_4$ is the most frequent in the estimated PMEM and also likely so in the given data. The horizontal axis of the disconnectivity graph does not bear any meaning; different local minima are horizontally spread out only for visualization purposes.
We also observe in Fig~\ref{fig:pipeline}(e) that the branch of $\bm{\alpha}_1$ and that of $\bm{\alpha}_2$ joins at $\overline{E}_{1, 2} = -1.2$, which is also consistent with Fig~\ref{fig:hypercube-with-paths}(b). In this manner, the disconnectivity graph represents all the local minima with their energy and the minimal height of the hill that any trajectory between each pair of local minima has to overcome.

The disconnectivity graph shown in Fig~\ref{fig:pipeline}(e) suggests grouping of the six local minima into three groups, one composed of $\bm{\alpha}_1$, $\bm{\alpha}_2$, and $\bm{\alpha}_3$, one composed solely of $\bm{\alpha}_4$, and the other composed of $\bm{\alpha}_5$ and $\bm{\alpha}_6$. Alternatively, one may want to avoid grouping $\bm{\alpha}_1$ together with $\bm{\alpha}_2$ and $\bm{\alpha}_3$. One obtains such a grouping by cutting the disconnectivity graph into subtrees by drawing an arbitrary horizontal line into the disconnectivity graph (e.g., at $E = -1.1$, as shown by the dotted line in Fig~\ref{fig:pipeline}(e)). For example, the authors of \cite{Watanabe2014NatComm} similarly grouped 10 local minima into three groups (shown in Fig~1(e) of \cite{Watanabe2014NatComm}). This procedure is the same as determination of the number of clusters in hierarchical clustering and necessarily arbitrary. Many studies relied on visual inspections or arbitrary criteria. We will explain a more principled method of grouping together some local minima into one major state in Section~\ref{sub:randomization}.

\subsection{Basin graph}

\begin{figure}[t]
\begin{center}
\includegraphics[width=14cm]{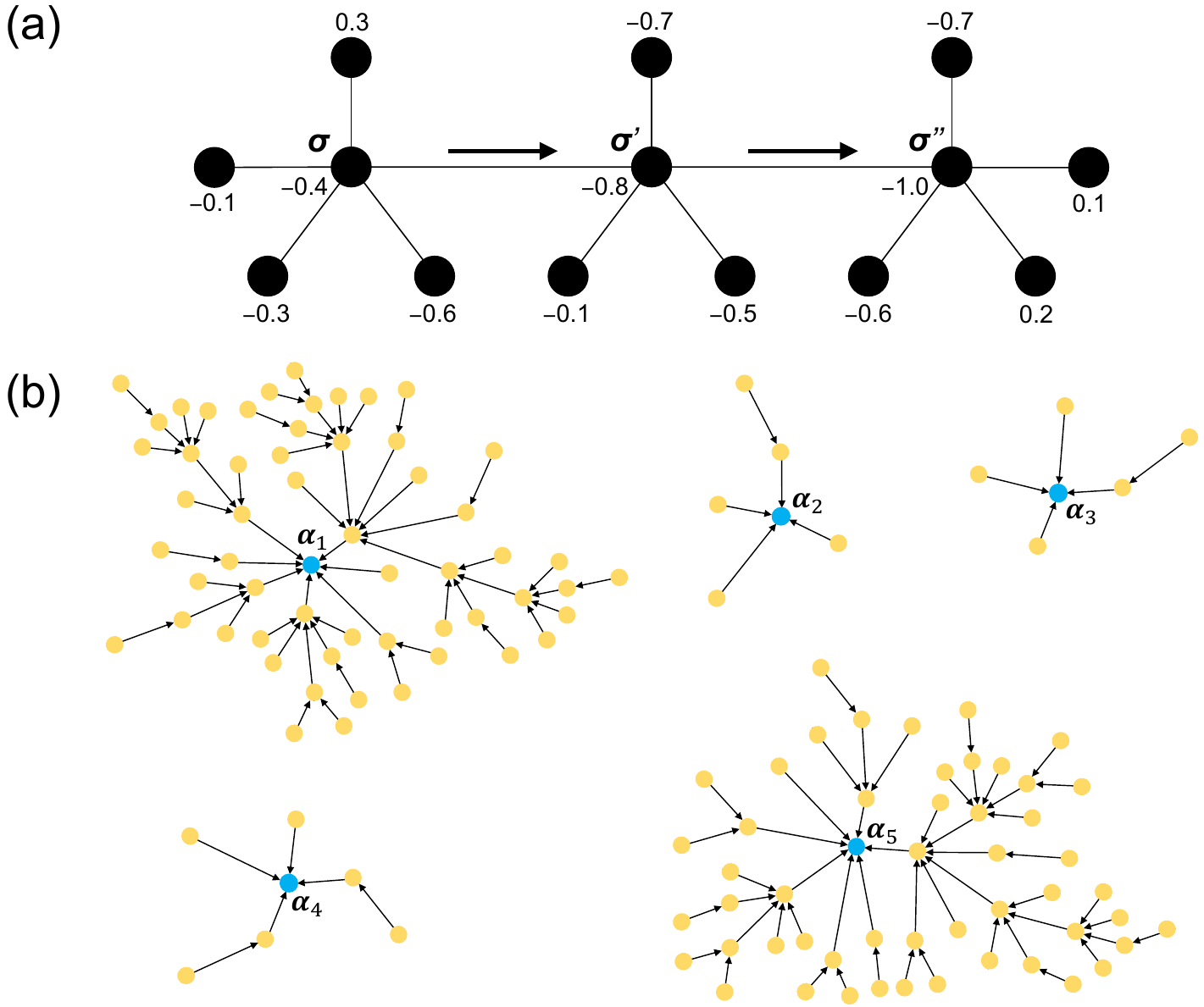}
\caption{Basin graph. (a) The steepest descent path from $\bm{\sigma}$ to its associated local minimum $\bm{\sigma}''$. A node represents an activity pattern. The numbers attached to each activity pattern represents its energy value. We set $N=5$. In fact, $\bm{\sigma}$ and $\bm{\sigma}''$ share two neighbors because they are distance 2 apart from each other. One such neighbor is $\bm{\sigma}'$. The other shared neighbor is an activity pattern with $E = -0.6$. For expository purposes, we showed this second neighbor as separate nodes in this figure. (b) An empirical basin graph with $N=7$ corresponding to the disconnectivity graph shown in Fig~\ref{fig:disconnectivity-graph-MSC}(a).}
\label{fig:basin}
\end{center}
\end{figure}

The disconnectivity graph only contains the local minimum activity patterns. How should we interpret activity patterns that are not local minima? The low-dimensional schematics shown in Figs~\ref{fig:1dim-schem} and \ref{fig:2dim-schem} suggest that activity patterns that are not local minima belong to the basin of a local minimum. In other words, if we release a ball at the position of an arbitrary activity pattern, $\bm{\sigma}$, on the energy landscape, and if the ball only goes downhill (which is contrary to the actual stochastic dynamics implied by the PMEM though), the ball eventually reaches one of the local minimum $\bm{\alpha}$ and stops there. In this case, we say that $\bm{\sigma}$ belongs to the basin of $\bm{\alpha}$. In reality, visual understanding, such as in Figs~\ref{fig:1dim-schem} and \ref{fig:2dim-schem}, is not easy to obtain because the network of activity patterns is a hypercube. However, the definition of the basin remains the same. We can tell
the basin to which an arbitrary activity pattern, $\bm{\sigma}$, belongs as follows. We start by identifying the
neighbor $\bm{\sigma}'$ of $\bm{\sigma}$ that attains the lowest energy among all the $N$ neighbors of $\bm{\sigma}$ (see Fig~\ref{fig:basin}(a) for a schematic). Unless $\bm{\sigma}$ is itself a local minimum, $\bm{\sigma}'$ has a lower energy than $\bm{\sigma}$, which we assume here. If there is a tie, we break it arbitrarily to set $\bm{\sigma}'$. The direction from $\bm{\sigma}$ to $\bm{\sigma}'$ is the steepest descent direction from $\bm{\sigma}$. Therefore, we move from $\bm{\sigma}$ and $\bm{\sigma}'$ in search of the bottom of the basin. Then, at $\bm{\sigma}'$, we similarly identify the neighboring activity pattern having the lowest energy, denoted by $\bm{\sigma}''$.
If $\bm{\sigma}''$'s energy is lower than $\bm{\sigma}'$'s energy, then we move to $\bm{\sigma}''$. Otherwise, $\bm{\sigma}'$ has an energy that is lower than all its neighbors such that $\bm{\sigma}'$ is a local minimum. In the latter case, we stop the journey at $\bm{\sigma}'$ and conclude that $\bm{\sigma}$ belongs to the basin of $\bm{\sigma}'$. If this is not the case, we repeat the procedure until we reach a local minimum, and we call that $\bm{\sigma}$ belongs to the basin of the finally visited activity pattern, which is a local minimum. In Fig~\ref{fig:basin}(a), the steepest descent path starting at $\bm{\sigma}$ stops at $\bm{\sigma}''$. Therefore, $\bm{\sigma}''$ is the local minimum associated with $\bm{\sigma}$. In this manner, we can identify to which basin each activity pattern belongs.

What we call the basin graph tells us the basin of each activity pattern and more; see Fig~\ref{fig:pipeline}(f) for a schematic and Fig~\ref{fig:basin}(b) for a real example. A node of the basin graph is an activity pattern. Therefore, there are $2^N$ nodes; in
Figs~\ref{fig:pipeline}(f) and \ref{fig:basin}(b), there are $2^5 = 32$ and $2^7=128$ nodes, respectively. 
The edge of the basin graph represents the steepest descent move from an activity pattern to its neighbor.
Differently from the network of activity patterns, which is also a network among the $2^N$ activity pattern nodes, the edge of the basin graph is directed, and each node has at most one out-going edge. If a node is not a local minimum, it has one out-going edge (i.e., the node's out-degree is equal to 1); see the orange nodes in Figs~\ref{fig:pipeline}(f) and \ref{fig:basin}(b). If a node is a local minimum, it does not have any out-going edge; see the blue nodes in the same figures. Any node has between 0 to $N$ in-coming edges; its number is called the in-degree of the node. The basin graph is composed of so-called weakly connected components (i.e., islands), each of which contains just one local minimum. This fact aids the interpretation of the basin graph.
The number of nodes in the connected components is equal to the size of the basin of the corresponding local minimum. Comparison between Figs~\ref{fig:pipeline}(e) and \ref{fig:pipeline}(f) and between Figs~\ref{fig:disconnectivity-graph-MSC}(a) and \ref{fig:basin}(b) suggests that an local minimum with a smaller energy value (i..e, therefore, high frequency of appearance) tends to have a large basin. However, while this association is often found in empirical data, it is not guaranteed.

Some activity patterns may be located near the border between different basins. Such an activity pattern roughly, but not exactly, corresponds to those on the top of a hill separating two basins (e.g., the local maximum in Fig~\ref{fig:1dim-schem} or a dotted line in Fig~\ref{fig:2dim-schem}). Precisely speaking, if activity pattern $\bm{\sigma}$ belonging to the basin of $\bm{\alpha}_i$ has a neighboring activity pattern $\bm{\sigma}'$ belonging to the basin of a different local minimum, then we say that $\bm{\sigma}$ and $\bm{\sigma}'$ are at the boundary between the two basins.
 
\subsection{Features of energy landscapes\label{sub:features}}

The disconnectivity graph and basin graph are visualizations of the energy landscape. They are useful for us to intuitively understand the dynamics generating the multivariate time series data and differences between different data or between different groups of data (e.g., between a group of healthy individuals and a group of those with disease) in terms of the energy landscape. However, we need to quantify such differences to draw conclusions. We refer to such a quantity as a feature of the energy landscape. A primary usage of a feature is to look for association between the feature and a property of interest in the research domain. For example, a schizophrenia researcher may be interested in seeking correlation between the severity of schizophrenia and a feature of the energy landscape. As another example, one may want to measure the effect of treatment in terms of the change in the value of a feature, especially if the feature is known to correlate with the severity of the disease.

A feature is also useful for assessing reliability of the ELA. If one measures a multivariate time series data of length $t_{\max}$ twice from the same system (e.g., same person in two separate days), run an ELA, and calculate a feature from the two energy landscapes, the value of the feature should be roughly the same between the two energy landscapes. Otherwise, one cannot rely on the ELA. Such a test, which should be done statistically, also requires a feature of energy landscape. Popular features of energy landscapes are as follows:

\begin{itemize}

\item
The number of local minima \cite{Watanabe2014FrontNeuroinfo, Kang2017Neuroimage}, denoted by $n_{\text{m}}$. Because a local minimum may not be significant in the sense that
it is separated from a different local minimum with a small energy barrier (see Section~\ref{sub:randomization} for a procedure to remove insignificant local minima), one may want to only count major local minima, i.e., significantly deep local minima.

\item The frequency with which a local minimum $\bm{\alpha}$ appears in the data \cite{Kang2017Neuroimage, Ezaki2018HumanBrainMapping}. It is the number of times $t$ at which $\bm{\sigma}(t) = \bm{\alpha}$, divided by $t_{\max}$.
Because the dynamics on the energy landscape are noisy by definition, it is more common to calculate the frequency with which the basin of a local minimum or that of a major local minimum is visited \cite{Watanabe2014NatComm, Kang2017Neuroimage, WatanabeRees2017NatComm, KangJeong2021HumanBrainMapping, Regonia2021FrontPsychiatry, Watanabe2021Elife, FanLiHuang2022NeuroimageClinical}.

\item The basin size \cite{Watanabe2014FrontNeuroinfo, Watanabe2014NatComm, Ashourvan2017Neuroimage, Kang2017Neuroimage, WatanabeRees2017NatComm}. It is the number of activity patterns belonging to the basin of a local minimum $\bm{\alpha}$, that can be divided by $2^N$ depending on the preference. It is equal to the number of nodes in the connected component of the basin graph to which $\bm{\alpha}$ belongs. The basin size tends to correlate with $E(\bm{\alpha})$ \cite{Watanabe2014FrontNeuroinfo} and the frequency of appearance of the basin \cite{Watanabe2021Elife}.

\item The energy barrier between local minima $\bm{\alpha}_i$ and $\bm{\alpha}_j$. It can be quantified by $\min\{ \overline{E}_{i, j} - E(\bm{\alpha}_i), \overline{E}_{i, j} - E(\bm{\alpha}_j) \}$, where we recall that $\overline{E}_{i, j}$ is the smallest energy value that any path between $\bm{\alpha}_i$ and $\bm{\alpha}_j$ needs to overcome \cite{Watanabe2014FrontNeuroinfo, Ashourvan2017Neuroimage, Kang2017Neuroimage}. The energy barrier quantifies the difficulty with which the system's state transits from $\bm{\alpha}_i$ to $\bm{\alpha}_j$ or vice versa. If the energy barrier is high, the transition occurs at a low rate in both directions.
Sometimes, the unsymmetrized, or directed, version of the energy barrier, i.e.,
$\overline{E}_{i, j} - E(\bm{\alpha}_i)$ or $\overline{E}_{i, j} - E(\bm{\alpha}_j)$, but not the minimum of the two, is used
\cite{Ashourvan2017Neuroimage, Watanabe2021Elife}. See Section~\ref{sub:random-walk} for quantitative relationships between the height of the energy barrier and the state transition rate.
 
A variant of this feature is an average over different local minima. In \cite{Khanra2024EurJNeurosci}, the authors used the average of branch length over all the major local minima in the pruned disconnectivity graph. By the pruned disconnectivity graph, we mean one created by sequentially removing insignificant local minima, to keep only the major local minima (see Section~\ref{sub:randomization} for the procedure). The branch length for local minimum $\bm{\alpha}_i$ is $\min_{j \text{ such that } j \neq i} \{ \overline{E}_{i, j} - E(\bm{\alpha}_i) \}$, where $\overline{E}_{i, j}$ now refers to the smallest energy at which the $i$th and $j$th branches of the pruned disconnectivity graph merge. For example, in Fig~\ref{fig:pipeline}(e), 
suppose that the pruning process discards $\bm{\alpha}_2$, $\bm{\alpha}_3$, and $\bm{\alpha}_6$, leaving $\bm{\alpha}_1$, $\bm{\alpha}_4$, and $\bm{\alpha}_5$ as major local minima. This pruning is equivalent to setting the threshold energy to the one shown by the dotted horizontal line in the figure to define three major states, each of which is represented by a major local minimum. Then, the branch length for $\bm{\alpha}_1$,
$\bm{\alpha}_4$, and $\bm{\alpha}_5$ are equal to
$\overline{E}_{1,4} - E(\bm{\alpha}_1)$, $\overline{E}_{1,4} - E(\bm{\alpha}_4)$, and $\overline{E}_{1,5} - E(\bm{\alpha}_5)$, respectively.
The last one is equal to $\overline{E}_{4,5} - E(\bm{\alpha}_5)$ because $\bm{\alpha}_1$ and $\bm{\alpha}_4$ merge earlier than $\bm{\alpha}_5$ does as we climb up the disconnectivity graph from a low energy value.
The average branch length used in \cite{Khanra2024EurJNeurosci} is the average of these three branch lengths and represents how distinct the different major local minima are from each other.

\item The frequency of transitions between basins \cite{Watanabe2014NatComm, WatanabeRees2017NatComm, Ashourvan2017Neuroimage, Ezaki2018HumanBrainMapping}. This is the
number of transitions from the basin of local minimum $\bm{\alpha}_i$ to that of local minimum $\bm{\alpha}_j$. To explain the idea, we note that the activity pattern at each time $t \in \{ 1, \ldots, t_{\max} \}$, i.e., $\bm{\sigma}(t)$, belongs to the basin of an $i$th local minimum, which we write as $\bm{s}(t) = i $, where each $\bm{s}(t) \in \{ 1, \ldots, n_{\text{m}} \}$. Then, the time series of the basin is given by $\{ \bm{s}(1), \ldots, \bm{s}(t_{\max}) \}$. The normalized frequency of transitions from the $i$th basin (i.e., basin of local minimum $\bm{\alpha}_i$) to the $j$th basin is the count of times with $\bm{s}(t) = i$ and $\bm{s}(t+1) = j$ over $t \in \{1, \ldots, t_{\max}-1 \}$, divided by $t_{\max}-1$. Another way to normalize the count is to divide it by the number of times at which $\bm{s}(t) = i$ over $t \in \{1, \ldots, t_{\max}-1 \}$. The latter quantity is equal to the empirical probability of transitions from $\bm{s}(t) = i$ to $\bm{s}(t+1) = j$ conditioned that $\bm{s}(t) = i$.

There are some variants of the frequency of transitions. 
First, one sometimes measures the number of transitions from the local minimum $\bm{\alpha}_i$, not from its basin, to $\bm{\alpha}_j$ \cite{Ezaki2018HumanBrainMapping, Kondo2022FrontNeurosci}.
A second variant is the frequency of indirect transitions from $i$ (either the local minimum or its basin) to $j$ through $k$ \cite{Watanabe2014NatComm, WatanabeRees2017NatComm, Watanabe2021Elife, FanLiHuang2022NeuroimageClinical, LiAnZhou2023FrontNeurosci, Watanabe2023Eneuro}.
In this count, we exclude the transitions from $i$ to $j$ that do not go through $k$. We call this type of transitions ``indirect''. Those that do not go through another local minimum or its basin are called direct transitions.
Third, one may want to count the number of iterations between two major local minima \cite{Watanabe2014NatComm}. This quantity is basically the sum of the number of transitions from $i$ to $j$ and that from $j$ to $i$.

\item Average dwell time of an $i$th basin \cite{Ashourvan2017Neuroimage, Kang2017Neuroimage}. This is the duration for which $\bm{\sigma}(t)$ resides in the $i$th basin (i.e., $\bm{s}(t) = i$) before leaving it. Because the dwell time is generally different for different visits to the $i$th basin in a single data set, we average the dwell time of the $i$th basin over the number of the visits to the $i$th basin. In one report, the average dwell time was strongly correlated with the basin size \cite{Ashourvan2017Neuroimage}.

\item The average number of time steps for moving from the $i$th basin to the $j$th basin to another, forming a path \cite{Watanabe2014NatComm}.

\item Various quantities that can be directly calculated from the estimated $J_{ij}$ or $h_i$ values. For example, a definition of the node strength (i.e., weighted degree) of the $i$th variable is $\sum_{j=1}^N | J_{ij} |$ \cite{Kang2017Neuroimage}. This class of features only requires the estimated PMEM, not further steps of the ELA.

\end{itemize}

One often needs to pool data from many different individual data, such as different participants in the experiment, and then estimate a single group PMEM with a high accuracy. Even in this situation, one can measure some of these features for each individual despite that only one energy landscape for the entire group of individuals is available. One can do this by using the group energy landscape and still computing these features for time series $\{ \bm{x}(1), \ldots, \bm{x}(t_{\max}) \}$ for each individual.
For example, one can calculate the dwell time of the $i$th basin for a single individual because the group energy landscape gives $\bm{\alpha}_i$ and its basin that are common to all the individuals.
In contrast, in the same situation, one cannot use features that are directly measured for the estimated energy landscape, not for
$\{ \bm{x}(1), \ldots, \bm{x}(t_{\max}) \}$ for each individual. Such features include $n_{\text{m}}$ and the energy barrier.

\section{Further methods}

In this section, we discuss advanced ELA methods, their backgrounds, and validation methods. 

\subsection{Phase diagram methods\label{sub:phase-diagram}}

An advantage of carrying out ELA using the Ising model is that
one can potentially benefit from statistical mechanical theory of the Ising spin-system model, which is a mature research field. The notions of local minima of energy and energy barrier have been investigated in spin-system theory for decades. One can also attempt to use other knowledge of spin-system theory for analyzing and interpreting multivariate time series data. One such method is to determine the phase of the multivariate time series data \cite{Ezaki2020CommunBiol}.

\begin{figure}[t]
\begin{center}
\includegraphics[width=7cm]{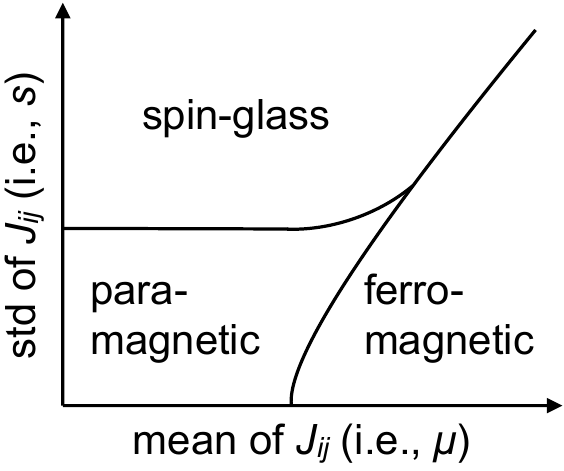}
\caption{Schematic of the phase diagram of the SK model. A more common parameter in physics literature is the temperature. However, in this figure, we chose the average and standard deviation of $J_{ij}$ because their interpretation is easier in our context.}
\label{fig:SK-phase-diagram}
\end{center}
\end{figure}

The statistical mechanical theory exploited here is that of prototypical spin systems, such as the 
Sherrington-Kirkpatrick (SK) model \cite{FischerHertz1991book}.  
The SK model is a type of Ising model and defined by drawing each $J_{ij}$ $(= J_{ji}; 1\leq i\neq j\leq N)$  independently from the Gaussian distribution with a given mean and standard deviation.
The SK model, as well as other spin system models, shows different phases depending on the parameter values (see Fig~\ref{fig:SK-phase-diagram} for a schematic).
In the paramagnetic phase, the spins (i.e., $\sigma_i$s) tend to be random and orderless in the sense that
the mean activity 
\begin{equation}
m = \frac{1}{N} \sum_{i=1}^N \langle \sigma_i \rangle_{\text{model}}
\end{equation}
and the so-called spin-glass order parameter defined by
\begin{equation}
q = \frac{1}{N} \sum_{i=1}^N \left(\langle \sigma_i \rangle_{\text{model}} \right)^2
\end{equation}
are both close to 0. Theoretically, $m$ and $q$ converge to $0$ as $N\to\infty$. The spin-glass order parameter $q$ represents the degree of local (i.e., for each variable) magnetization, or the local stickiness of the activity. In an energy landscape, the paramagnetic phase corresponds to a situation in which local minima are shallow and not distinguishable from those existing in completely random spin systems. Therefore, if we run a significance test for the local minima in this situation (see Section~\ref{sub:randomization} for a method), local minima in the paramagnetic phase should be classified to be insignificant.

In the ferromagnetic phase, the spins strongly tend to align in one direction, let say, $\sigma_i = 1$. It is defined by $m \neq 0$ and $q > 0$.
For example, $m>0$ if a majority of $\sigma_i$s tends to be $1$. A strongest case of ferromagnetic phase corresponds to an energy landscape in which $\bm{\sigma} = (1, \ldots, 1)$ is by far the deepest local (and global) minimum. However, in ELA, we usually impose that $\langle \sigma_i \rangle_{\text{data}}$ is close to $0$ by choosing a threshold on $x_i(t)$ as such. Therefore, $\langle \sigma_i \rangle_{\text{model}}$ is also expected to be close to $0$, and $m$ substantially far from $0$ is not likely to occur. Instead, in practice, we often have an energy landscape in which $\bm{\sigma} =  (1, \ldots, 1)$ and $\bm{\sigma} = (-1, \ldots, -1)$ are the only most dominant local minima
\cite{Ezaki2018HumanBrainMapping, Kondo2022FrontNeurosci, LiAnZhou2023FrontNeurosci, Miyata2024PsychiatryClinicalNeurosci}.
%
%

The other phase is the spin-glass phase, which has attracted a lot of statistical physics research.
The spin-glass phase corresponds to
rugged energy landscapes (and therefore, complex state-transition dynamics) \cite{FischerHertz1991book} and chaotic dynamics \cite{Bray1987PhysRevLett, Rizzo2003PhysRevLett, Aspelmeier2008PhysRevLett}.
It is defined by $m=0$ and $q>0$ (in the $N\to\infty$ limit).
In our ELA, the obtained energy landscape would have relatively many local minima whose branches are not too short to be ignorable.

Criticality, or phase transition, i.e., the boundary between (typically) the paramagnetic phase and the spin-glass phase in a parameter space 
(see Fig~\ref{fig:SK-phase-diagram}), is of special interest due to rich dynamical behavior of the spin system near there. On these grounds, Ezaki et al.\,developed a data-driven method to locate a given multivariate time series data as a point in the phase diagram to tell which phase the given data resides in, and quantify how close the data is from the boundary with a different phase \cite{Ezaki2020CommunBiol}.

The first step of the method is to fit the Ising model to the given data, as in the ordinary ELA.
Because the above discussion holds true in principle in the limit of $N\to\infty$, we need to estimate a relatively large Ising model. If $N$ is small, phase transition lines would be blurred. We used $N=264$ in \cite{Ezaki2020CommunBiol}. Then, we need to give up an exact estimation method and resort to an approximate model estimation method such as pseudo-likelihood maximization. A large $N$ leads to a low accuracy of fit, which is an issue to be investigated in future research.

Next, to draw a phase diagram from the given multivariate time series data, we fix $\bm h \equiv (h_1, \ldots, h_N)$ at that for the estimated PMEM by following a convention \cite{FischerHertz1991book, Tkacik2015PNAS}.
Under this constraint, we vary the mean of $\bm{J} \equiv (J_{12}, J_{13}, \ldots, J_{N-1, N})$, denoted by
$\mu$, and the standard deviation of $\bm{J}$, denoted by $s$, to linearly transform $\bm{J}$ as follows:
\begin{equation}
J_{ij} = (\hat{J}_{ij} - \hat{\mu}) \frac{s}{\hat{s}}  + {\mu},
\label{eq:scale-J}
\end{equation}
where $\hat{\mu}$, for example, is the value of $\mu$ for the originally estimated PMEM. With Eq~\eqref{eq:scale-J}, if we arbitrarily choose the values of $\mu$ and $s$, then we obtain $\bm{J}$, which specifies one PMEM.
For each $\bm{J}$ specified by a $(\mu, s)$ pair, we perform Monte Carlo simulations to
produce $\{\bm{\sigma}(1), \bm{\sigma}(2), \ldots \}$ and calculate observables such as $m$ and $q$. In this manner, we can generate a phase diagram for each observable in terms of $\mu$ and $s$, as in Fig~\ref{fig:SK-phase-diagram}.

With this method, energy landscapes estimated for fMRI data from healthy human participants were in the paramagnetic phase but close to the boundary between the paramagnetic and spin-glass phases \cite{Ezaki2020CommunBiol}. It was also shown that the participants with higher intelligence quotient scores were, albeit weakly, closer to the phase boundary, consistent with the so-called critical brain hypothesis 
\cite{Ezaki2020CommunBiol}.

\subsection{Validation methods and good practices\label{sub:validation}}

Because ELA fits a model (i.e., PMEM) to data, it is imperative to validate the methods on the data. The accuracy of fit introduced in Section~\ref{sub:accuracy-fit} is one validation method. In this section, we explain other validation methods.

\subsubsection{Threshold variation}

The choice of the threshold, $\theta_i$, for reducing each $i$th time series $\{x_i(1), \ldots, x_i(t_{\max}) \}$ into a binarized time series $\{ \sigma_i(1), \ldots, \sigma_i(t_{\max}) \}$ is an arbitrary parameter. We recommend that one tests the robustness of the results, such as the accuracy of fit, the disconnectivity graph at the level of major local minima, and association between a feature of the an energy landscape and, e.g., a cognitive score for a human participant in the experiment, for a range of values of  $\theta_i$. For example,
in some studies, it was verified that the accuracy of fit remained large for a range of $\theta_i$, $i \in \{1, \ldots, N \}$, in which the frequency of $\sigma_i = 1$ (as opposed to $\sigma_i=-1$) varied between approximately 30\% and 70\% \cite{Watanabe2013NatComm, Watanabe2014Neuroimage, Ezaki2018HumanBrainMapping}.
%

\subsubsection{Randomization\label{sub:randomization}}

In Fig~\ref{fig:pipeline}, local minimum $\bm{\alpha}_2$ is shallow in the sense that the energy barrier between $\bm{\alpha}_2$ and $\bm{\alpha}_3$ is small. Therefore, the dwell time in $\bm{\alpha}_2$ would be small. Therefore, although $\bm{\alpha}_2$ is technically a local minimum, it may be better to ignore it in practice. Only the local minima that have sufficiently high energy barriers with others are trustable.
We already discussed this topic; by choosing a threshold energy level, shown by the dotted line in Fig~\ref{fig:pipeline}(e), one can classify the local minima into groups. However, how should we choose the threshold?
How can one tell whether or not a local minimum is too shallow to be seriously considered? This decision has been rather arbitrary in most studies using the ELA. The root of the problem is basically the same as an often asked problem for clustering algorithms for data: how to determine the number of clusters?

For ELA, we developed a statistical test to tell which local minima are significant~\cite{Khanra2024EurJNeurosci}.
Specifically, to generate a completely random disconnectivity graph to be constrasted with the disconnectivity graph from the given data, we generate a completely random multivariate time series of the same size (i.e., same $N$ and $t_{\max}$) as that of the original time series data by assigning $-1$ or $1$ with probability $0.5$ each, independently to each of $\sigma_i(t)$, $i \in \{1, \ldots, N \}$, $t \in \{1, \ldots, t_{\max} \}$. For the generated random multivariate time series, we estimate the PMEM and then the disconnectivity graph. A completely random time series should not have any meaningful structure including local minima. Therefore, we regard that the branch length in the generated random disconnectivity graph (see Section~\ref{sub:features} for the definition of the branch length)
for any $\bm{\alpha}$ is meaningless. To operationalize this idea, we record the maximal branch length among all the local minima in the random disconnectivity graph.

We compute the maximal branch length for a sufficiently large number, $c$, of completely random multivariate time series data. We set $c=10^2$ in \cite{Khanra2024EurJNeurosci} and use $c=10^3$ in the following running example in this section.
We denote the mean and standard deviation of the maximal branch length over the $c$ random disconnectivity graphs by $\langle L \rangle$ and $\text{std}(L)$, respectively.
Then, we remove the local minimum with the shortest branch length in the original disconnectivity graph as being insignificant if its branch is shorter than $\langle L \rangle + 2 \times \text{std}(L)$. 
For example, Fig~\ref{fig:disconnectivity-graph-MSC}(b) shows the disconnectivity graph for a random multivariate binary time series having the same $N$ ($=7$) and $t_{\max}$ ($= 3,272$) as those for the original data whose disconnectivity graph is shown in Fig~\ref{fig:disconnectivity-graph-MSC}(a). Due to the randomization, the disconnectivity graph in Fig~\ref{fig:disconnectivity-graph-MSC}(b) has shorter branch lengths than the disconnectivity graph in Fig~\ref{fig:disconnectivity-graph-MSC}(a) in general. The maximal branch length in Fig~\ref{fig:disconnectivity-graph-MSC}(b) is that of $\bm{\alpha}^{\prime}_3$ and equal to $L = 0.125$.
%
%
Based on $c = 10^3$ random disconnectivity graphs, we obtained
$\langle L \rangle = 0.0902$ and $\text{std}(L) = 0.0471$. Therefore, $\langle L \rangle + 2 \times \text{std}(L) = 0.184$. 
%
%
The shortest branch of the empirical disconnectivity graph shown in Fig~\ref{fig:disconnectivity-graph-MSC}(a), which is
for $\bm{\alpha}_3$, is shorter than $0.184$. Therefore, we regard $\bm{\alpha}_3$ as an insignificant local minimum and prune it away.

If we have removed a local minimum according to this procedure, then we recompute the branch length of the local minima affected by the removal. In Fig~\ref{fig:disconnectivity-graph-MSC}(a), removal of the branch of $\bm{\alpha}_3$ does not affect the branch length of any other local minima, especially because $\bm{\alpha}_4$ and $\bm{\alpha}_5$ merge before they merge $\bm{\alpha}_3$. Therefore, there is no need to recompute any branch length. After the possible recomputation, we repeat the above procedure. In other words, if the shortest branch is shorter than $\langle L \rangle + 2\times \text{std}(L)$, we remove the branch and recompute the affected branch lengths. For example, in Fig~\ref{fig:disconnectivity-graph-MSC}(a), we next remove the branch of $\bm{\alpha}_2$ because it is shorter than the threshold (i.e., $0.184$). This removal changes the branch length for $\bm{\alpha}_1$ from 0.814
%
%
to 1.082.
%
%
Next, we remove the branch of $\bm{\alpha}_4$, which increases the branch length for $\bm{\alpha}_5$ from $0.761$
%
%
to 0.902.
%

We repeat the procedure until the branches of all the surviving local minima are longer than the threshold in the pruned disconnectivity graph.
We refer to the surviving local minima as major local minima.
For example, for Fig~\ref{fig:disconnectivity-graph-MSC}(a), the final, pruned discoonnectivity graph, shown in 
Fig~\ref{fig:disconnectivity-graph-MSC}(c), only consists of $\bm{\alpha}_1$ and $\bm{\alpha}_5$, which are the two major local minima.
This pruning algorithm also tells us how to associate each removed local minimum and its basin to
a major state. In Fig~\ref{fig:disconnectivity-graph-MSC}(a), $\bm{\alpha}_1$, $\bm{\alpha}_2$, and their basins constitute a major state represented by $\bm{\alpha}_1$. Similarly, $\bm{\alpha}_3$, $\bm{\alpha}_4$, $\bm{\alpha}_5$, and their basins constitute a major state represented by $\bm{\alpha}_5$.

There are two remarks.
First, the threshold $\langle L \rangle + 2 \times \text{std}(L)$ is arbitrary and probably too generous. We test the significance of the branch length at most $n_{\text{m}}$ times, where we recall that $n_{\text{m}}$ is the number of local minima. Therefore, one may want to set a more conservative threshold, such as $\langle L \rangle + 3 \times \text{std}(L)$ or one that explicitly carries out Bonferroni correction,
to avoid a potential multiple comparison problem. 
Second, one may want to use a different method to generate random multivariate time series of the same size (i.e., $N\times t_{\max}$). In fact, it is a common practice to randomize the given time series, creating surrogate time series, then measure a quantity to characterize time series, $q$, and compare the value of $q$ with that for the original time series to assess whether the $q$ value for the original time series is statistically meaningful \cite{Theiler1992PhysicaD, Schreiber2000PhysicaD}. We can implement such a procedure here, too. However, it should be noted that time stamp shuffling (i.e., randomly permute the time stamps of $\{\bm{\sigma}(1), \ldots,
\bm{\sigma}(t_{\max}) \}$ does not work because the PMEM ignores the time order of the data.
The PMEM estimated for the original data and that estimated for the data after time stamp shuffling are exactly the same.

\subsubsection{Random-walk simulations\label{sub:random-walk}}

Once the PMEM is estimated, one can generate random-walk trajectories on the hypercube to emulate the dynamics of the system's state with a high time resolution (because each step of the walk is only to the nearest neighbor on the hypercube by definition) and for arbitrary long time \cite{Watanabe2014NatComm, WatanabeRees2017NatComm, Ashourvan2017Neuroimage}. We consider discrete-time random walks for simplicity and construct them such that the moves from activity pattern $\bm{\sigma}$ to $\bm{\sigma}'$ and vice versa respect the detailed balance condition, i.e.,
\begin{equation}
\frac{T_{\bm{\sigma} \to \bm{\sigma}'}} {T_{\bm{\sigma}' \to \bm{\sigma}}} = e^{E(\bm{\sigma}) - E(\bm{\sigma}')},
\label{eq:detailed-balance}
\end{equation}
where $T_{\bm{\sigma} \to \bm{\sigma}'}$ is the transition probability of the random walk from $\bm{\sigma}$ to $\bm{\sigma}'$. This detailed balance condition is a quantitative formulation of the already made claim that the movement of the state uphill on the energy landscape is not much likely to occur, especially when the energy barrier to be overcome to go to a different activity pattern is high. In this way, moves from a high energy activity pattern to a low energy one are more likely than vice versa. However, the random walk is a stochastic process, so moves from a lower energy activity pattern to a higher one do occur with some probability. Note that Eq~\eqref{eq:detailed-balance} is consistent with the stationary density given by Eq~\eqref{eq:P(V)}.

Random-walk simulations are useful for validation of the ELA. For example, we can calculate
the transition probability from a major state represented by a major local minimum $\bm{\alpha}$ to an adjacent major state represented by major local minimum $\bm{\alpha}'$, denoted by $T_{\bm{\alpha} \to \bm{\alpha}'}^{(\text{basin})}$, from both random-walk simulations and the given data. If these two numbers are drastically different, the ELA may not be a suitable method for the data. Such a discrepancy, if it exists, may be because the transition probability in the given data is more complex than what the detailed balance condition dictates. For example, if the probability of the move $\bm{\sigma} \to \bm{\sigma}'$ depends on where the trajectory comes from to $\bm{\sigma}$, i.e., the last activity pattern before $\bm{\sigma}$, it should not be fitted by the PMEM in principle. This is because the PMEM, hence ELA, does not assume such a dependence. Alternatively, if the given data is too short, we may not observe sufficiently many transitions from the basin of $\bm{\alpha}$ to that of $\bm{\alpha}'$ such that the statistical fluctuation in $T_{\bm{\alpha} \to \bm{\alpha}'}^{(\text{basin})}$ estimated for the given data is too large.
 
A practical method to run random-walk simulations is a Markov chain Monte Carlo method with the Metropolis-Hastings algorithm \cite{Walter2015PhysicaA} (used in, e.g., \cite{Watanabe2014NatComm}).
In each discrete time step, the random walker selects a neighbor, $\bm{\sigma}'$, of the currently visited node, $\bm{\sigma}$, with equal probability $1/N$. The walker actually moves from $\bm{\sigma}$ to $\bm{\sigma}'$ with probability $T_{\bm{\sigma} \to \bm{\sigma}'} = \min \{1, e^{E(\bm{\sigma}) - E(\bm{\sigma}')} \}$. Otherwise, the walker does not move.
We remark that, if one prefers, main features of the random walk such as the dwell time and transition probabilities, including those between basins, can be theoretically calculated without running simulations \cite{Masuda2017PhysRep}. This is because the complete transition probability matrix of the random walk on the hypercube with $2^N$ nodes is available once the PMEM is estimated.

\subsubsection{Test-retest reliability}

If ELA is reliable, the energy landscape constructed from the data collected in one day should be similar to that on another day unless the system has sufficiently changed between the two days. Otherwise, we cannot trust the constructed energy landscapes even if the accuracy of the fit values are high. We recently developed a statistical test of test-retest reliability for experimental designs where data are collected in multiple sessions from each participant of the experiment and over multiple participants \cite{Khanra2024EurJNeurosci}. The main idea is to compare two energy landscapes constructed from two disjoint sets of sessions from the same participant, called the within-participant comparison, and two energy landscapes constructed from different participants, called the between-participant comparison. If ELA is reliable, the within-participant comparison should yield a smaller discrepancy score than the between-participant comparison, which we operationalized. Because one session is often too short to yield a high accuracy of fit, we built the above procedure to also work when one estimates each energy landscape from data pooled across sessions. In other words, a within-participant comparison is made between two energy landscapes from the same participant, with each energy landscape being built from $\tilde{m}$ sessions from the same participant. Then, we need at least $2\tilde{m}$ sessions of data from the same participant. The obtained within-participant comparison result is to be contrasted with the between-participant comparison, which consists in comparing the energy landscape constructed from $\tilde{m}$ sessions from $\tilde{m}$ participants, one session from each participant, and that constructed from $\tilde{m}$ sessions from other $\tilde{m}$ participants.

\subsection{Challenges\label{sub:challenges}}

We recognize two main difficulties in the original ELA.

First, ELA is data-hungry. If we aim at maintaining a high accuracy of fit of the PMEM, we roughly need to increase the data length, $t_{\max}$, twice as we add one variable (i.e., as we increase $N$ by $1$). Therefore, $t_{\max}$ needed increases exponentially in terms of $N$, i.e.,  in proportion to $2^N$ \cite{Ezaki2017PhilTransRSocA}. Then, if we seriously take the accuracy, one cannot go beyond approximately $N=10$ or $N=15$, depending on $t_{\max}$ of the data, while ignoring the fitting accuracy to conduct a large-scale ELA (e.g., see Section~\ref{sub:phase-diagram}) is a methodological choice.

For this reason, in neuroimaging data analysis, in which the ELA has been most intensively used, it is customary to pool data from different participants in the experiment to enhance the size of the data. A typical approach is to pool data across participants in a control group and run an ELA, estimating just one energy landscape. Then, we do the same for a set of participants in a treatment group (such as those receiving the same diagnosis) and compare the results between the two groups. This procedure ameliorates the data length issue at a cost: the data from different participants in either group may not be as homogeneous as that from a single participant. If the participants within a group are too heterogeneous, then estimating a common ELA from the pooled data may not be justified.

Note that, when pooling data, we may need to use different threshold values, $\theta_i$, for binarizing $x_i(t)$ for different participants. For example, assume that one participant, denoted by $p_1$, tends to have a large $x_i(t)$. We also assume that a different participant classified to the same group, denoted by $p_2$, tends to have a small $x_i(t)$. Then, if we use $\theta_i$ shared by everybody in the group, $x_i(t)$ from $p_1$ tends to yield $\sigma_i(t)=1$ and $x_i(t)$ from $p_2$ tends to yield $\sigma_i(t) = -1$. However, this is not what ELA intends to do; ELA intends to analyze the system's dynamics for each participant under the assumption that the dynamics are similar among the participants in the same group. A common method to evade this situation is to set $\theta_i$ separately for different participants such that, for example, the fraction of $\sigma_i(t) = -1$ and that of $\sigma_i(t) = 1$ are approximately the same (i.e., 50\%) for each participant (e.g., \cite{Ezaki2018HumanBrainMapping, Khanra2024EurJNeurosci}).

Second, binarization implies loss of information contained in the original multivariate time series data. If the original time series data for each variable, $\{x_i(1), \ldots, x_i(t_{\max}) \}$, shows bimodality, binarization is better justified. However, bimodal distributions of a time series are not frequently found in many domains. Despite the apparent loss of information by binarization, the ELA has been successful in many studies. The reason for this is unknown.

In the next two sections, we introduce two methods that expand the original ELA to address these problems.

\subsection{Variational Bayes approximation method\label{sub:Bayesian}}

Consider a collection of multivariate time series data from different participants in the same experiment. We assume that these participants belong to the same group (e.g., people with a particular psychiatric disorder). It is typical that the data from a single participant is not long enough to be able to estimate the PMEM with a high accuracy. If we pool the data across all the participants in the group, the combined data may be sufficiently long, but we may suffer from heterogeneity in the data among the different participants, as we discussed in Section~\ref{sub:challenges}. Because the PMEM implicitly assumes that observed data points, no matter whether they come from the same participant or different participants, are drawn independently from the same distribution, such a heterogeneity, which may be better modeled by a participant-dependent distribution of data, is undesirable. In this situation, a variational Bayes approximation method \cite{KangJeong2021HumanBrainMapping, JeongKang2021Neuroimage} enables us to estimate the PMEM, and hence the energy landscape, separately for each participant while benefiting from some commonality that the participants in the same group have. Core steps of this Bayesian method are as follows.

Assume that there are $D$ multivariate time series data (such as those from $D$ participants) belonging to the group.
%
%
%
This method uses a prior distribution of the PMEM parameters specific to each time series data in the group (referred to as ``series'' for short in the remainder of this section), denoted by
\begin{equation}
\bm{\theta}_{D'} = (\theta_{D'1}, \ldots, \theta_{D'M}) \equiv (h_1, h_2, \ldots, h_N, J_{12}, J_{13}, \ldots, J_{N-1,N}) \in \mathbb{R}^M
\end{equation}
for the $D'$th series, where $D' \in \{ 1, \ldots, D \}$, and $M \equiv N(N+1)/2$ is the number of parameters of the PMEM. We set the prior distribution of
\begin{equation}
\Theta=[\bm{\theta}_1, \ldots, \bm{\theta}_D]
\end{equation}
to
\begin{equation}
p(\Theta | \bm{\eta}, \bm{\tilde{\alpha}})=\prod_{D'=1}^D \prod_{M'=1}^M p(\theta_{D' M'} | \mathcal{N} (\eta_{M'}, 1/\tilde{\alpha}_{M'})),
\label{eq:prior-all-data}
\end{equation}
where $p(x | \mathcal{N} (\mu, \sigma^2))$ represents the probability density of $x$ obeying the one-dimensional normal distribution with mean $\mu$ and variance $\sigma^2$. Here, $\bm \eta=(\eta_1, \ldots, \eta_M)^\top \in \mathbb R^M$ is the prior mean vector, $\bm{\tilde{\alpha}} = (\tilde{\alpha}_1, \ldots, \tilde{\alpha}_M)^\top \in \mathbb R_+^M$ is the prior precision vector, and ${}^{\top}$ represents transposition. Equation~\eqref{eq:prior-all-data} implies that the data from the $D$ series are mutually independent.

It is intractable to derive the posterior distribution of $\Theta$ because the normal distribution is not a conjugate prior for the Boltzmann distribution. Therefore, we use a variational approximation to the posterior distribution~\cite{Bishop2006book} using the normal distribution as follows:
\begin{equation}
q(\Theta |\mathcal{S}, \bm{\eta}, \bm{\tilde{\alpha}}) = \prod_{D' =1}^D \prod_{M'=1}^M p(\theta_{D' M'} | \mathcal{N} (\mathbb \mu_{D' M'}, 1/\beta_{D' M'})),
\end{equation}
where $q$ is the posterior distribution, and $\mathcal{S}$ is the set of the data from all the $D$ series.

We denote the posterior mean vector and the posterior precision vector for the $D'$th series by $\bm{\mu}_{D'}=(\mu_{D' 1}, \ldots, \mu_{D' M})^\top \in \mathbb{R}^M$ and $\bm{\beta}_{D'}=(\beta_{D' 1}, \ldots, \beta_{D' M})^\top \in \mathbb{R}_+^M$, respectively.
One obtains the variational approximate solution for $q$ by optimizing the evidence lower bound (ELBO), which is also called the free energy~\cite{KangJeong2021HumanBrainMapping, JeongKang2021Neuroimage}. By maximizing the free energy with respect to $q$, we calculate the posterior quantities for each $D'$th series, $\bm{\mu}_{D'}$ and $\bm{\beta}_{D'}$, from the prior mean vector $\bm{\eta} = (\eta_1, \ldots, \eta_M)^{\top}$ and the prior precision vector $\bm{\tilde{\alpha}} = (\tilde{\alpha}_1, \ldots, \tilde{\alpha}_M)^{\top}$ as follows:
\begin{eqnarray}
\bm{\mu}_{D'} &=& \bm{\eta} + t_{\max} \mathbb A_{\bm{\eta}, \bm{\tilde{\alpha}}}^{-1}(\langle \bm{\bar{\sigma}} \rangle_{\text{data}, D'} - \langle \bm{\bar{\sigma}} \rangle_{\bm{\eta}}),
\label{mu_D'}\\
\bm{\beta}_{D'} &=& \bm{\tilde{\alpha}} + t_{\max} \bm{c}_{\bm{\eta}},
\label{beta_D'}
\end{eqnarray}
where 
\begin{eqnarray}
\mathbb A_{\bm{\eta}, \bm{\tilde{\alpha}}} &=& \text{diag}(\bm{\tilde{\alpha}})+ t_{\max}  \text{C}_{\bm{\eta}},
\end{eqnarray}
is an $M\times M$ matrix, and $\rm{diag}(\cdot)$ represents the diagonal matrix whose entries are given by the arguments.
In Eq~\eqref{mu_D'},
\begin{equation}
\langle \bm{\bar{\sigma}} \rangle_{\text{data}, D'} \equiv (\langle \sigma_1 \rangle_{\text{data}, D'}, \ldots, \langle \sigma_N\rangle_{\text{data}, D'}, \langle \sigma_1 \sigma_2 \rangle_{\text{data}, D'}, \langle \sigma_1 \sigma_3 \rangle_{\text{data}, D'}, \ldots, \langle \sigma_{N-1} \sigma_N \rangle_{\text{data}, D'})^{\top}
\end{equation}
is the vector composed of the mean activity of the individual variables and the mean pairwise joint activities of the data;
vector $\langle \bm{\bar{\sigma}} \rangle_{\bm{\eta}}$ is the model mean of
\begin{equation}
\bm{\bar{\sigma}} \equiv  (\sigma_1, \ldots, \sigma_N, \sigma_1 \sigma_2, \sigma_1 \sigma_3, \ldots, \sigma_{N-1} \sigma_N)^{\top}
\end{equation}
when the PMEM parameters $(h_1, \ldots, h_N, J_{12}, J_{13}, \ldots, J_{N-1,N})$ are given by $\bm{\eta}$;
matrix $\rm C_{\bm{\eta}}$ is the $M \times M$ covariance matrix of $\bm{\bar{\sigma}}$ when the PMEM parameters are $\bm{\eta}$.  
In Eq~\eqref{beta_D'}, $\bm{c}_{\bm{\eta}}$ is the vector composed of the diagonal entries of $\rm C_{\bm{\eta}}$.
In other words, the $i$th entry of $\bm{c}_{\bm{\eta}}$ is the variance of the $i$th entry of $\bm{\bar{\sigma}}$ given PMEM parameters $\bm{\eta}$.

Now, we fix $q$ and maximize the free energy with respect to $\bm{\eta}$ and $\bm{\tilde{\alpha}}$ to obtain the updating rules for $\bm{\eta}$ and $\bm{\tilde{\alpha}}$ as follows:
\begin{eqnarray}
\eta_{M'} &=& \frac{1}{D}\sum_{D'=1}^D \mathbb \mu_{D' M'} \label{eta}, \\
\tilde{\alpha}_{M'} &=& \left[ \frac{1}{D} \sum_{D'=1}^D \left\{(\mu_{D' M'}-\eta_{M'})^2 + \frac{1}{\beta_{D' M'}} \right\} \right]^{-1}, \label{alpha}
\end{eqnarray}
where $M' \in \{1, \ldots, M\}$.

Thus, we have updated the posterior distribution $\theta_{D' M'} \sim \mathcal{N}(\mu_{D' M'}, 1/\beta_{D' M'})$, $D' \in \{1, \ldots, D\}$, $M' \in \{1, \ldots, M\}$ using the prior distribution $\theta_{D' M'} \sim \mathcal{N}(\eta_{M'}, 1/\tilde{\alpha}_{M'})$, and then updated the prior distribution using the obtained posterior distribution. These operations constitute one cycle of the algorithm. The steps of the variational Bayes approximation method are as follows:
\begin{enumerate}
\item Initialize the hyperparameters by independently drawing each $\eta_{M'}$ (with $M' \in \{1, \ldots, M\}$) from the normal distribution with mean $0$ and standard deviation $0.1$. We also set the first $N$ entries of the prior precision vector $\bm{\tilde{\alpha}}$, corresponding to $h_i$, $i\in \{1, \ldots, N\}$, to $6$, and
set the remaining $M-N$ entries of $\bm{\tilde{\alpha}}$, corresponding to $J_{ij}$ (with $1\le i < j \le N$), to $30$. 
These values are used in \cite{KangJeong2021HumanBrainMapping, JeongKang2021Neuroimage} and their accompanying code.

\item Calculate the posterior mean vector, $\bm{\mu}_{D'}$, and the posterior precision vector, $\bm{\beta}_{D'}$, for each $D' \in \{ 1, \ldots, D\}$ using Eqs~\eqref{mu_D'} and \eqref{beta_D'}.

\item Update the prior mean vector, $\bm{\eta}$, and the prior precision vector, $\bm{\alpha}$, using Eqs~\eqref{eta} and \eqref{alpha}.

\item If $\left\lvert \frac{\rm{ELBO}(\rm{iter})}{\rm{ELBO}(\rm{iter}-1)}-1\right\rvert < 10^{-8}$, we stop iterating steps 2 and 3. Otherwise, we return to step 2. Here, $\rm{ELBO}(\rm{iter})$ represents the ELBO value after `iter' iterations.
\end{enumerate}

\subsection{Clustering and Markovian approaches\label{sub:clustering}}

An attempt to mitigate possible loss of information by binarizing the originally continuous signal is to dichotomize the original signal into a larger number of discrete values. However, this idea does not help much. For example, using ternary activity for each variable (i.e., $\sigma_i(t) \in \{ -1, 0, 1 \}$) did not enhance the accuracy of fit in one study \cite{Watanabe2013NatComm}. This is presumably because, by introducing a larger number of discrete activity levels for each variable, we have less samples (i.e., time points in the case of time series) in each activity pattern. If there are three activity levels per variable, there are $3^N$ possible activity patterns as opposed to $2^N$. If many of the $3^N$ possible activity patterns are never visited by the data or their counts are close to $0$, the estimation of the corresponding PMEM may not be accurate.

Another strategy is to fit a different model than the PMEM to try to realize the same goal as the ELA. The main goals of ELA are to identify a small number of major states from the multivariate time series data in a data-driven manner and to interpret the time series as a journey from one identified major state to another and further.

One simple alternative to the ELA along this idea is to cluster the $t_{\max}$ data points in $\mathbb{R}^N$ using any clustering algorithm and regard each cluster as a major state \cite{LiuChangDuyn2013FrontSystNeurosci, LiuZhangChang2018Neuroimage, Ezaki2021EurJNeurosci, Paakki2021BrainBehav, Islam2024BmcNeurosci}. In the EEG data analysis for the brain, such clustering analysis, or more sophisticated versions of this, is known as microstate anaylsis \cite{Khanna2015NeurosciBiobehavRev, Michel2018Neuroimage}. Application of various conventional clustering methods used in microstate anaylsis to fMRI data has shown that, in terms of various features, the test-retest reliability of conventional clustering methods is not particularly behind the ELA on the same data sets
\cite{Islam2024BmcNeurosci}. This result is encouraging because the clustering methods avoid dichotomization of data. Furthermore, one can carry out many (albeit not all) measurements of energy landscapes also for clustering analysis. For example, one can count the transitions from one cluster to another and compute the dwell time in each cluster. If transitions from cluster $C$ to cluster $C'$ and vice versa are considerably rarer than typical cluster-to-clulster transitions, one can make an analogue to ELA to regard that $C$ and $C'$ are separated by a high energy barrier.

Hidden Markov models (HMMs) applied to multivariate time series share the goal of capturing the given data as transitory dynamics among discrete states. They see applications to, e.g., neuroimaging data \cite{Baker2014Elife, Ezaki2021EurJNeurosci, Tait2022Neuroimage-toolbox}. The rate of transition from one state to another particular state and the dwell time in each state, for example, are features the original time series data as in the case of the conventional clustering algorithms. A key difference between the HMM and the conventional clustering is that the HMM assumes that the state transition rate depends on the currently visited state, whereas the conventional clustering approach does not. The HMM explicitly estimates the state transition rate for each pair of states. In contrast, in the clustering approach, one only estimates discrete clusters, or states, without looking at dynamics of the time series data. Then, one calculates the state transition rate as the normalized number of the state transitions (e.g., from state $i$ to $j$) present in the data.

The ELA, or PMEM, is closer to the conventional clustering approach than to the HMM in this aspect. In estimating the PMEM from data, only the equilibrium distribution of the data is fitted, similar to conventional clustering. With the PMEM, one does not fit the state transition rate from each particular state to another. The state transition rates in the case of ELA are determined by the detailed balance condition (i.e., Eq~\eqref{eq:detailed-balance}) and the structure of the estimated disconnectivity graph.

It may sound that the HMM is better than conventional clustering and the ELA because the HMM allows the explicit estimation of the individual state-transition rates dependent on the current state. However, the HMM usually has more parameters to be estimated, due to the state-dependent state-transition rate parameters, at least than clustering analyses. Therefore, if the given data are relatively short or the number of variables is large, it may be better to give up the HMM to run for conventional clustering or ELA. In addition, poor temporal resolution and strong autocorrelation of the data may also make the estimation of state-transition rates in the HMM difficult.

\section{Software\label{sec:software}}

We are aware of the following open-resource code packages for ELA and the related methods introduced in this article.

The energy landscape analysis toolbox (ELAT), accompanying a review paper \cite{Ezaki2017PhilTransRSocA}, is a MATLAB package with a graphical user interface \cite{Ezaki2017PhilTransRSocA-code-ELAT}. It can estimate the PMEM from data, compute the accuracy of fit, basin graphs in two and three dimensions (the three-dimensional version shows the energy value of each activity pattern on the vertical axis), disconnectivity graph, list of activity patterns at each local minimum (each being an $N$-dimensional binary vector, as in the gray-white vectors shown in Fig~\ref{fig:disconnectivity-graph-MSC}), time series of the activity pattern, time series of the basin ID to which the activity pattern at each time belongs, and some features of the constructed energy landscape.
The code accompanying an MEG study \cite{Krzeminski2020NetwNeurosci} is provided in MATLAB \cite{Krzeminski2020NetwNeurosci-code}. It uses ELAT and provides analyses complementary to ELAT such as calculation of the best threshold for binarization and a boostraping for statistical analysis of energy values.
A Python implementation of ELAT is available \cite{Yonezawa2024IntJMolSci-code} as an accompaniment of 
\cite{Yonezawa2024IntJMolSci}.
Code for carrying out ELA is also available by other authors in R \cite{Suzuki2024rELA} and Mathematica \cite{Suzuki2024ela-mathematica}, as an accompaniment of \cite{Suzuki2021EcolMonographs}.
The code accompanying an EEG study
\cite{Klepl2022IeeeJBiomedHealthInfo} is also written R and capable of estimating the PMEM by pseudo-likelihood maximization and carrying out ELA~\cite{Klepl2022IeeeJBiomedHealthInfo-software}.

The code accompanying \cite{JeongKang2021Neuroimage} is written in R and provides PMEM estimation by the standard maximum likelihood and Bayesian methods~\cite{JeongKang2021Neuroimage-software}.
The code accompanying \cite{Khanra2024EurJNeurosci} is provided mostly in MATLAB and partly in R
\cite{Khanra2024-code}. It allows us to estimate the standard PMEM, calculate significant local minima, run the variational Bayes method, and run test-retest reliability analyses.
The code accompanying \cite{Ezaki2020CommunBiol} is provided in MATLAB \cite{Ezaki2020CommunBiol-code}. It also allows us to estimate the standard PMEM and provides part of code for the phase diagram methods explained in Section~\ref{sub:phase-diagram}.
Code for clustering-based estimation of discrete states as an analogue of ELA (see Section~\ref{sub:clustering}) is available
\cite{Islam2024BmcNeurosci-code} as an accompaniment of \cite{Islam2024BmcNeurosci}.

\section{Applications\label{sec:appl}}

In this section, we showcase some applications of ELA.

In our first application, we studied energy landscapes of fMRI data obtained from healthy humans undergoing a visual bistable perception task \cite{Watanabe2014NatComm}. We used $N=7$ biologically informed locations in the brain, called regions of interest (ROI), as variables for constructing the PMEM. We constructed a single energy landscape from the fMRI data during the task. For the data pooled across the participants, we found ten local maxima and three heuristically determined major local maxima. In general, by inspecting the activity pattern at each major local minimum, one can see which ROI is relatively active (i.e., $\sigma_i(t) = 1$) or inactive (i.e., $\sigma_i(t) = -1$), based on which one can biologically characterize the major local minima. One major state was an aggregation of five local minima and their basins, and we named this major state the frontal-area state because the ROIs in the frontal area of the brain were active at the major local minimum defining this major state. Likewise, the visual-area state was composed of two other local minima and their basins. The visual-area state is complementary to the frontal-area state in that the ROIs active in the former are inactive in the latter and vice versa. The third major state, named the intermediate state, contained three local minima. These three major states remained the same for the PMEM separately estimated for the data from each single participant.
In the bistable visual perception task, the participants were asked to watch a set of moving dots on the screen, which allows bistable perception (i.e., either of the two directions of the apparently moving sphere composed of dots). The participants reported whenever their subjective perception spontaneously switched from one direction to the other. The main result was that participants with a larger basin size of the frontal-area state tended to have a shorter duration of the percept (i.e., how long the participant perceives one direction of motion of the visual stimulus before it subjectively switches to the other direction). In contrast, a larger basin size of the visual-area state tended to yield a longer duration of the percept. In this manner, we found an association between behavior (i.e., the median duration of the percept) and brain dynamics revealed by ELA (i.e., basin sizes), as schematically shown in Fig~\ref{fig:pipeline}(g).

ELA also succeeded in identifying atypical brain mechanisms underlying prevalent neuropsychiatric conditions. One study on autism spectrum disorder (ASD) using fMRI data found that both high-functioning ASD adults and control individuals exhibited similar energy landscapes with the same two major states, each composed of two local minima that were the same between the ASD and control groups. The ASD adults showed atypically reduced indirect transitions between the two major states that went through one of the two local minima not belonging to a major state \cite{WatanabeRees2017NatComm}. In particular, this specific transition frequency was so closely correlated with the severity of the ASD symptoms that it could be used for accurate diagnosis of autism. In a similar manner, another resting-state fMRI study showed that attention deficit hyperactivity disorder (ADHD) is characterized by atypical increases in overall transition probabilities between basins of local minima \cite{Watanabe2023Eneuro}.

Multistability is often present in the composition of multispecies ecosystems, and different compositions may have different impacts on biodiversity, resistance to infection, and other functions. Suzuki et al.\,was among the first to apply the ELA based on the Ising model to outside neuroscience \cite{Suzuki2021EcolMonographs}. They analyzed gut microbiota data obtained from feces of the same six mice over the course of 72 weeks. They collected 18 data points per mouse, totaling $18 \times 6 = 108$ data points in each week in which the measurement was made. They tracked the changes in the energy landscape on $N=8$ variables, which were the binarized abundance of bacterial species, as the mice aged. They found that the energy at the local minima gradually decreased (i.e., the local minima became more frequent) over the age and that the energy barrier between the two local minima with the smallest energy also decreased, making the transition between them easier for older mice.

\section{Outlook}

We have provided a tutorial of ELA based on the Ising model, with an emphasis on concepts and steps that are often confusing to users. We conclude this article by stating some possible directions for future research.

The problem of the data length, $t_{\max}$, is inherent in the ELA because $t_{\max}$ required for a given accuracy seems to scale with $2^N$ \cite{Ezaki2017PhilTransRSocA}. Given this situation, effectiveness of the Bayesian methods (Section~\ref{sub:Bayesian}), which tactically combines the strength of pooling of the data across many time series to earn an effective $t_{\max}$ and enable estimation of the Ising model for each time series, should be further investigated. Possible research includes applications to other data sets and developments of other variants of Bayes approximation methods. In a broader context, transfer learning for the PMEM \cite{Yosinski2014Nips, ZenMyTan2020PhysRevE} may help mitigate the data paucity problem. In particular, pooling of different data sets (e.g., similar data obtained from experiments carried out by different research groups) as well as pooling of different time series data within one data set may worth pursuing.

We provided a list of existing applications of the ELA in the introduction section. While ELA has been progressively used in other domains, most applications so far have been to neuroscience data, especially fMRI data, presumably reflecting the history of the method. However, we emphasize that the ELA is completely domain-general and can be applicable to any multivariate time series data as long as there are at least about seven variables (i.e., $N\ge 7$) and $t_{\max}$ is large enough to justify the choice of $N$. Because of the binarization step, ELA may be particularly useful when each variable shows a bimodal distribution over time, such as expression of some genes
\cite{Gormley2008BmcBioinfo, Eldar2010Nature, Mason2011BmcGenom, Aqil2024biorxiv}. In fact, the fitting of the Ising model to neuronal spiking data \cite{Schneidman2006Nature, Shlens2006JNeurosci, Tang2008JNeurosci, YuHuangSinger2008CerebralCortex, Yeh2010Entropy, Mora2011JStatPhys}, which preceded the ELA based on the Ising model, benefited from the observation that a neuron's state is approximately binary in the sense that it either spikes or it does not at any given time. With this remark in mind, applications of ELA to unforeseen types of multivariate time series data would be exciting.

In Section~\ref{sub:phase-diagram}, we introduced a phase diagram analysis, which enables us to qualitatively tell properties of the given data set. If the system generating the observed time series data is changing on a slow timescale, which may be due to the environmental change (i.e., gradual pollution of an ecosystem) or the internal changes of the system (e.g., aging of the participant), the energy landscape of the system may transit from one phase to another (e.g, from the paramagnetic to spin-glass phase) on a slow timescale. A related challenge is to find early warning signals for large regime shifts of the system, which may occur as a result of bifurcations of the dynamical system underlying the data or for different reasons. Popular early warning signals include the sample variance and the decay rate of the sample autocorrelation function obtained from the observed time series, which would increase as the system approaches a bifurcation point \cite{Scheffer2009Nature, Scheffer2012Science, Boettiger2013TheorEcol, Dakos2015PhilTransRSocB,Southall2021JRSocInterface}. A large regime shift that one wants to anticipate may correspond to a mass extinction of species in an ecosystem or a rapid progression of a disease in a human individual. From an energy landscape perspective, these methods usually assume that a large regime shift occurs when a local minimum in which the system's state is currently sitting is lost via the diminishment of the energy barrier between the currently visited basin and a neighboring basin; see Fig~1 in \cite{Scheffer2009Nature}, Figs 2 and 3 in \cite{ Scheffer2012Science}, and Box 1 in \cite{ Dakos2015PhilTransRSocB} for schematics. Then, the regime shift corresponds to the movement of the system's state from the current basin to the neighboring basin. Cross-fertilization of ELA and early warning signal research may be interesting.

ELA is not only a powerful tool for inferring and interpreting brain state dynamics but also for modifying neural dynamics and relevant behaviors. A recent study has developed a non-invasive closed-loop neural stimulation system using transcranial magnetic stimulation (TMS) and demonstrated that one can modify the height of a specific energy barrier and control the dynamics of perceptual consciousness in humans \cite{Watanabe2021Elife}.
Also see \cite{Gupta2023Sensors}. Such intervention methods should be useful when we know desirable versus undesirable energy landscapes or when we can measure the effect of changing the energy landscape. While the TMS is considered to locally change $h_i$ and hence $\sigma_i$ for nodes $i$ in the region of the brain affected by the TMS, other types of intervention, especially those that changes the connectivity between specific pairs of nodes, may be possible depending on the system \cite{ParkKang2021FrontComputNeurosci}.
Simulation studies are expected to be useful for understanding specificity of intervention \cite{Kang2017Neuroimage, KangPae2019PlosOne, ParkKang2021FrontComputNeurosci}. Further studies along this line, including applications outside neuroimaging, are warranted.

\section*{Acknowledgements}

We acknowledge Pitambar Khanra for technical information.
%

\end{document}